\documentclass[twocolumn,amsfonts,showpacs,superscriptaddress,amsmath,amssymb,aps, longbibliography,prd]{revtex4-2}

\usepackage{physics}
\usepackage{graphicx}
\usepackage{amsthm}

\newcommand{\prlsection}[1]{{\em {#1}.---~}}

\newtheorem{result}{Result}

\newcommand{\upd}[1]{^\mathrm{#1}}
\newcommand{\ind}[1]{_\mathrm{#1}}
\DeclareMathOperator{\sgn}{\mathrm{sgn}}

\def\eqrefsm#1{(#1)}

\begin{document}
\newcommand{\titleinfo}{A new quadrature for the generalized hydrodynamics equation\\ and absence of shocks in the Lieb-Liniger model}
\title{\titleinfo}
\author{Friedrich H\"ubner}
\affiliation{Department of Mathematics, King’s College London, Strand, London WC2R 2LS, U.K.}
\author{Benjamin Doyon}
\affiliation
{Department of Mathematics, King’s College London, Strand, London WC2R 2LS, U.K.}
%
%
\begin{abstract}
In conventional fluids, it is well known that Euler-scale equations are plagued by ambiguities and instabilities. Smooth initial conditions may develop shocks, and weak solutions, such as for domain wall initial conditions (the paradigmatic Riemann problem of hydrodynamics), are not unique. The absence of shock formation experimentally observed in quasi-one-dimensional cold-atomic gases, which are described by the Lieb-Liniger model, provides perhaps the strongest pointer to a modification of the hydrodynamic equation due to integrability. Generalised hydrodynamics (GHD) is the required hydrodynamic theory, taking into account the infinite number of conserved quantities afforded by integrability.  We provide a new quadrature for the GHD equation -- a solution in terms of a Banach fixed-point problem where time has been explicitly integrated. The quadrature is an efficient numerical solution tool; and it allows us, in the Lieb-Liniger model, to rigorously show that no shock may appear at all times, and, when combined with recent hydrodynamic fluctuation theories, to obtain new expressions for correlations in non-stationary states, establishing for the first time the presence of discontinuitites characteristic of the non-equilibrium dynamics.
\end{abstract}

\maketitle

\prlsection{Introduction} In recent years there has been an increased interest in studying one of the oldest and most powerful framework for many-body systems out of equilibrium: that of hydrodynamics. It captures the large space-time scale behaviour of local observables, and is built out of the local conservation laws admitted by the system, and the thermodynamic equations of state. While in most physical systems there are only a few local conservation laws (such as particle number, momentum and energy), one-dimensional {\em integrable systems} admit an infinite number. A new universality class for their hydrodynamic equations has been uncovered, referred to as generalised hydrodynamics (GHD)~\cite{PhysRevX.6.041065,bertini2016transport,el2003thermodynamic}. Most notably, GHD provides the relevant equation for the large-scale dynamics of quasi-one-dimensional cold atomic gases realised in experiments \cite{PhysRevLett.122.090601,10.21468/SciPostPhys.6.6.070,PhysRevLett.126.090602,doi:10.1126/science.abf0147} and described by the Lieb-Liniger model \cite{lieb1963}, as reviewed in \cite{Bouchoule_2022}. It is also the hydrodynamic theory underlying kinetic equations of soliton gases observed in optic fibres and water tanks \cite{PhysRevE.109.061001}. GHD applies to almost every known many-body integrable system (spin chains, field theories, classical and quantum gases, etc.) \cite{10.21468/SciPostPhysLectNotes.18,Bastianello_2022,ESSLER2023127572,doi:10.1142/13600}.

The Euler-scale dynamics of integrable models is described by a ``quasi-particle" density in phase space, $\rho(t,x,\lambda)$, which encodes the densities of all conservation laws, typically including the particle, momentum and energy densities. The variable $\lambda$ parametrises a {\em continuum} of hydrodynamic modes. The quasi-particle density has a simple physical definition: in the gas, take out a mesoscopic-length fluid cell element $[x,x+\dd x]$, put it in the vacuum, and let it expand for a long time. Then the quantity $\rho(t,x,\lambda)\dd x\dd\lambda$ is the number of particles (or more generally asymptotic objects, such as solitons, waves, etc.) observed with asymptotic momenta lying in $[\lambda,\lambda+\dd\lambda]$. This process is experimentally realisable \cite{doi:10.1126/science.abf0147}.

A drastic phenomenological difference between solutions to conventional hydrodynamic (CHD) equations for non-integrable systems, and GHD, is the appearance of shocks. Early attempts to explain experiments on quasi-one-dimensional cold atomic gases by using CHD failed dramatically because in CHD, typically, shocks develop in finite time. Yet experimentally, the gas appears to be spatially smooth at all times. In GHD, it was indeed observed that shocks do not develop \cite{PhysRevLett.119.195301}. In fact, finite-component reductions of GHD have the property of ``linearly degeneracy'' \cite{el2011kinetic,pavlov2012generalized}, known to prevent the appearance of shocks \cite{FERAPONTOV1991112,lindeg1,lindeg2}, and a continuum analogue of this property holds for GHD \cite{10.21468/SciPostPhysLectNotes.18}. Shocks encode physics beyond the Euler scale, and break uniqueness of solutions, requiring physically motivated entropy-production conditions \cite{bressan2000hyperbolic}. This is clearest in the Riemann problem of hydrodynamics: a domain-wall initial condition (``partitioning protocol'') is not a well-defined initial-value problem in CHD. By contrast GHD solutions appear to be well defined, composed of a continuum of contact singularities \cite{PhysRevX.6.041065}, which are entropy-preserving. The appearance of shocks and entropy production at the Euler scale is a fundamental aspect of the large-scale physics of conventional systems, and their observed absence, the starkest signature of integrability. Given that, it is important to show, beyond any doubt, that GHD indeed does not develop shocks and is well-posed even for the Riemann problem.

In this paper, we introduce a new solution method for the GHD equation, a ``quadrature'' presented as integral equations where the space-time variables have been explicitly integrated, which significantly improces on earlier quadratures \cite{DOYON2018570,PhysRevLett.119.220604}. The new quadrature has both practical and conceptual importance. First, it gives a new, numerically efficient {\em solution algorithm} for GHD, which, we believe, surpasses previous algorithms. Second, we explain how it gives rise to {\em a rigorous proof} (carried out in our companion paper~\cite{long}), in the Lieb-Liniger model, that smooth initial conditions stay smooth at all times -- hence no shocks develop --, and that weak solutions are unique and entropy-preserving -- hence contact singularities give the unique solution to the Riemann problem. Thus, we have established the  phenomenology of the GHD equation for cold atomic gases. Third, it also allows us to obtain new {\em explicit formula for large-scale correlation functions}. Going beyond previous studies~\cite{10.21468/SciPostPhys.5.5.054,10.21468/SciPostPhysCore.3.2.016,PhysRevLett.131.027101} we establish, from this, that the equal-time two-point correlation functions instantaneously develops a PT-breaking jump at equal points, by contrast to equilibrium correlations.

\prlsection{GHD in normal coordinates}
The GHD equation for single particle type (see the SM~\cite{SM} for more general framework) \cite{PhysRevX.6.041065,bertini2016transport,10.21468/SciPostPhysLectNotes.18,el2003thermodynamic},
\begin{align}
	\partial_t \rho(t,x,\lambda) + \partial_x (v\upd{eff}(t,x,\lambda)\rho(t,x,\lambda)) &= 0,\label{equ:GHD_conservation}
\end{align}
describes the evolution of the quasi-particle density $\rho(t,x,\lambda)$. The so called effective velocity $v\upd{eff}(t,x,\lambda)$ of the quasi-particles satisfies the  following self-consistency equation:
\begin{align}\label{equ:veff}
	&v\upd{eff}(t,x,\lambda) = v(\lambda)\\
	&+ \int\dd{\mu} \varphi(\lambda,\mu)\rho(t,x,\mu)(v\upd{eff}(t,x,\mu)-v\upd{eff}(t,x,\lambda)).\nonumber
\end{align}
\begin{figure}
	\centering
	\includegraphics{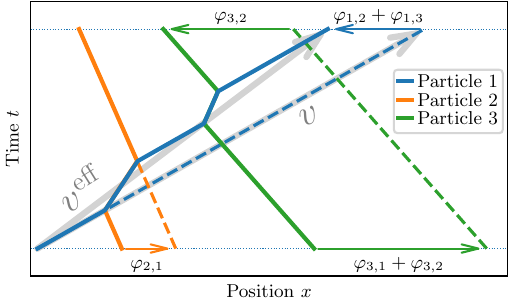}
	\vspace{-0.7cm}
	\caption{Illustration of the intuition behind GHD \cite{el2011kinetic,pavlov2012generalized,PhysRevLett.120.045301} (here using the so-called semi-classical Bethe models~\cite{PhysRevLett.132.251602}): The quasi-particles (solid lines) obtain a Wigner-time delay during scattering, leading to an effective position shift $\varphi_{ij}=\varphi(\lambda_i,\lambda_j)$. Therefore, the particles effectively move with a different velocity $v\upd{eff}\neq v$. The normal coordinate $\hat{x}$ (dashed lines) is obtained by adding the scattering shifts of all particles to their left to their position $x$, which in the continuum limit becomes \eqref{equ:x_hatx}. In these new coordinates the evolution is non-interacting with velocity $v(\lambda_i)$.\vspace{-0.5cm}}
	\label{fig:solitons}
\end{figure}
The intuition behind this equation is illustrated in Fig. \ref{fig:solitons}.
Each quasi-particle evolves with its own bare (or asymptotic) velocity $v(\lambda)$. During the scattering of a particle with momentum $\lambda$ with a particle with momentum $\mu$, effectively, the particle trajectory is displaced by a semi-classical spatial shift $\varphi(\lambda,\mu)$, similar to two-soliton scattering of integrable PDEs. The effective velocity is the bare velocity plus the contribution from the scattering with other particles, which is proportional to the scattering shift $\varphi(\lambda,\mu)$, the density $\rho(t,x,\mu)$ and the differences of effective velocities (this is because faster particles scatter with more particles in the same unit time). In the Lieb-Liniger model, with quantum Hamiltonian $H = \sum_i p_i^2/2 + (c/2)\sum_{i\neq j} \delta(x_i-x_j)$, one finds $v(\lambda) = \lambda$ and $\varphi(\lambda,\mu) = 2c/(c^2 + (\lambda-\mu)^2)$ \cite{lieb1963}. 

Equation \eqref{equ:GHD_conservation} may seem rather unwieldy: it is highly non-linear, and non-local in the momentum $\lambda$ variable, involving the solution to an integral equation. Finite-element numerics is possible (for state of the art solvers see the IFluid project~\cite{10.21468/SciPostPhys.8.3.041}), but there other, more efficient solution schemes.

In order to understand them, we recall the following fundamental property of GHD, established in \cite{DOYON2018570}, at the basis of its physics. Consider the rapidity- and state-dependent coordinate change $x\mapsto \hat x = \hat{X}_i(t,x,\lambda)$, with
\begin{align}
	\hat{X}(t,x,\lambda)= x + \int_{-\infty}^x\dd{y}\int\dd{\mu}\varphi(\lambda,\mu)\rho(t,y,\mu).\label{equ:x_hatx}
\end{align}
Then the quasi-particle density in the new coordinates,
\begin{align}
	\hat{\rho}(t,\hat{X}(t,x,\lambda),\lambda)\dd{\hat x}= \rho(t,x,\lambda)\dd{x},\label{equ:rho_hat_rho}
\end{align}
satisfies the {\em free-particle kinetic equation:}
\begin{align}
	\partial_t \hat{\rho}(t,\hat{x},\lambda) + v(\lambda)\partial_{\hat{x}}\hat{\rho}(t,\hat{x},\lambda) &= 0.\label{equ:GHD_hat_rho}
\end{align}

Physically, the ``normal coordinate" $\hat x$ is a position that takes into account the ``density of available space" (see Fig. \ref{fig:solitons}). Indeed, the quantity 
\begin{align}
    \textstyle2\pi\rho_{{\rm s}}(t,x,\lambda)=
\tfrac{{\rm d} \hat{X}(t,x,\lambda)}{x}= 1+\int\dd{\mu}\varphi(\lambda,\mu)\rho(t,x,\mu)\label{equ:rhos}
\end{align}
has the interpretation as a (non-negative) density of the space available for particles to travel through freely \cite{DOYON2018570} (the factor $2\pi$ is conventional). This can be seen similar to the metric in general relativity, which changes the physical space based on the amount of mass present. In the $\hat{x}$ space trajectories of particles are linear. This picture makes only sense if $\rho_{{\rm s}}(t,x,\lambda)>0$ is positive: there must remain a positive amount of space; thus $\hat X(t,x,\lambda)$ is monotonically increasing in $x$. The map \eqref{equ:x_hatx} is a generalisation of the ``excluded volume" technique used in the hard-rod gas \cite{Spohn1991} and extended and used in order contexts \cite{PhysRevE.104.044106, PhysRevE.104.064124, PhysRevLett.132.251602, doyon2023generalisedtbartdeformationsclassicalfree, Croydon2021}, and is fundamentally related to the factorised scattering property of integrable systems \cite{10.21468/SciPostPhysLectNotes.18}.

\prlsection{Advantage of quadratures} In \cite{DOYON2018570} this was used to find a quadrature for the GHD equations, where time $t$ appears only as external parameter. Indeed, Eq. \eqref{equ:GHD_hat_rho} is solved trivially, $\hat{\rho}(t,\hat{x},\lambda) = \hat{\rho}(0,\hat{x}-v(\lambda)t,\lambda)$; and $\hat{\rho}(t=0,\hat{x},\lambda)$ can be computed directly from $\rho(t=0,x,\lambda)$. Thus $\hat{\rho}(t,\hat{x},\lambda)$ is known at any time $t$ and one only need solve for $\rho(t,x,\lambda)$ in \eqref{equ:x_hatx} and \eqref{equ:rho_hat_rho}. An algorithm for this was proposed and illustrated in \cite{DOYON2018570},
and later used in \cite{10.21468/SciPostPhysCore.3.2.016}.

The advantages of quadratures of this type are clear: (1) in order to compute the solution at time $t$ one does not need to compute the solution at any other time, and thus numerical inaccuracies introduced by finite-time elements disappear; and (2) the computational effort to obtain the solution remains constant as time increases, unlike in finite-element schemes whose computational effort increases with $t$. Therefore, such iterative schemes are useful in particular for simulating long time dynamics, which is crucial for hydrodynamic phenomenology. However, the algorithm proposed in \cite{DOYON2018570} is relatively time-consuming and numerically delicate, involving whole-space numerical integrations. Because of this, in applications these advantages are lost.


\prlsection{The new quadrature}
In this work we present a powerful new quadrature. It allows us to solve both the conceptual and practical problems discussed above: to show, rigorously, the absence of shocks and the uniqueness of solutions at all times; and to find the solution directly at any fixed space-time point $(t,x)$, giving an efficient solution algorithm. The idea is to introduce height-fields
\begin{align}
     \Psi(t,x,\lambda) &= \textstyle\int_{-\infty}^x\dd{y}\rho(t,y,\lambda)&
    \hat{\Psi}(t,\hat{x},\lambda) &= \textstyle\int_{-\infty}^{\hat{x}}\dd{\hat{y}}\hat{\rho}(t,\hat{y},\lambda),
\end{align}
which satisfy the equations
\begin{align}
    \partial_t \Psi(t,x,\lambda) + v\upd{eff}(t,x,\lambda)\partial_x\Psi(t,x,\lambda) &= 0\label{equ:psi_equ}\\
    \partial_t \hat{\Psi}(t,\hat{x},\lambda) + v(\lambda)\partial_{\hat{x}}\Psi(t,\hat{x},\lambda) &= 0\label{equ:psi_hat_equ}.
\end{align}
Note that \eqref{equ:psi_equ} in the usual hydrodynamic terminology means that $\Psi(t,x,\lambda)$ is a Riemann invariant (there are many more such Riemann-invariants, see SM \cite{SM}). From \eqref{equ:psi_hat_equ} one can again solve explicitly $\hat{\Psi}(t,\hat{x},\lambda) = \hat{\Psi}(0,\hat{x}-v(\lambda)t,\lambda)$. Furthermore, one can show from \eqref{equ:rho_hat_rho} that $\hat{\Psi}(t,\hat{X}(t,x,\lambda),\lambda) = \Psi(t,x,\lambda)$. Using this and \eqref{equ:x_hatx} we find:
\begin{align}
    \Psi(t,x,\lambda)
    &= \hat{\Psi}\qty(0,x-v(\lambda)t+\int\dd{\mu}\varphi(\lambda,\mu)\Psi(t,x,\mu),\lambda)\label{equ:fixed_point_Psi}.
\end{align}
Alternatively we can plug this into \eqref{equ:x_hatx} and find
\begin{align}
    \hat{X}(t,x,\lambda) &= x + \int\dd{\mu}\varphi(\lambda,\mu) \hat{\Psi}\qty(0,\hat{X}(t,x,\mu)-v(\mu) t,\mu)\label{equ:fixed_point_hatX}.
\end{align}
\eqref{equ:fixed_point_Psi} and \eqref{equ:fixed_point_hatX} are ``self-consistent" fixed-point equations for $\Psi(t,x,\lambda)$ and $\hat{X}(t,x,\lambda)$ respectively. Observe that they fully decouple for different $t,x,$. Solution to such equations can be obtained for instance by fixed-point iteration, for which strong results on convergence are known (see below). Once a solution is found, one can either compute the solution from \eqref{equ:fixed_point_Psi} via $\rho(t,x,\lambda) = \partial_x\Psi(t,x,\lambda)$ or from \eqref{equ:fixed_point_hatX} by first computing the so-called occupation function
\begin{align}
    n(t,x,\lambda) = 2\pi\hat{\rho}(0,\hat{X}(t,x,\lambda)-v(\lambda)t,\lambda)\label{equ:n_rhohat}.
\end{align}
By taking a derivative of \eqref{equ:fixed_point_Psi} we find
\begin{align}
    \rho(t,x,\lambda) &= n(t,x,\lambda)\Big(1 + \int\tfrac{\dd{\mu}}{2\pi}\varphi(\lambda,\mu) \rho(t,x,\mu)\Big).\label{equ:rho_from_n}
\end{align}
This linear equation can either be solved again by fixed-point iteration, or by using a linear solver.
While taking the derivative might seem simpler, solving \eqref{equ:rho_from_n} has the advantage that it is also independent of $t$ and $x$. Therefore we conclude:

\begin{result}
    By solving first \eqref{equ:fixed_point_hatX} and then \eqref{equ:rho_from_n} one is therefore able to find the solution to the GHD equation directly at an arbitrary space-time point $t,x$. It is only required to solve self-consistency equations in the remaining degree of freedom $\lambda$, and these equations have clear convergence properties.
\end{result}

\begin{figure}
    \centering
    \includegraphics{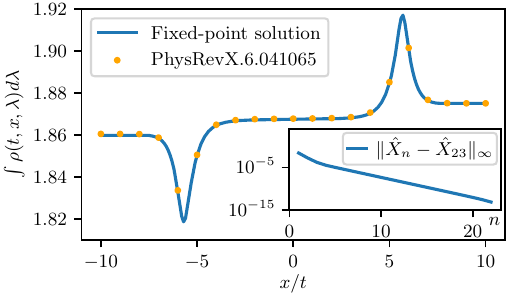}
    \vspace{-0.9cm}
    \caption{Momentum-integrated particle density profile of a partitioning protocol in the Lieb-Liniger model. The initial state and model parameters correspond to Fig. 8 in~\cite{PhysRevX.6.041065}. The result obtained via fixed-point iteration (solid line) of \eqref{equ:fixed_point_hatX} agrees with the solution given in~\cite{PhysRevX.6.041065} (dots). The inset shows the error of $\hat{X}(t=1,x=0,\lambda)$ after the $n$'th iteration compared to the final result approximated by taking $n=23$ steps; we see the exponential convergence.\vspace{-0.5cm}}
    \label{fig:numerics}
\end{figure}

We demonstrate that this new efficient algorithm works in practice by reproducing a simulation of a previous publication~\cite{PhysRevX.6.041065}, see Fig. \ref{fig:numerics}.

\prlsection{Absence of shocks, existence and uniqueness} We will now demonstrate on the example of the repulsive Lieb-Liniger model that the new fixed-point approach is extremely useful beyond providing a new efficient algorithm. We will show that its GHD equation \eqref{equ:GHD_conservation} has a unique solution, which furthermore does not develop shocks. 

For this we need the well-known fact that the quasi-particles of the Lieb-Liniger model have fermionic statistics, which means that $n(t,x,\lambda)<1$ (from the thermodynamic Bethe ansatz the occupation function can be interpreted as the average occupation of quantum numbers -- this can be maximum $1$ in fermionic systems). In principle $n(t,x,\lambda)=1$ is also possible, but we exclude this boundary case for now and discuss it in the end matter section \ref{sec:precise}. Using \eqref{equ:n_rhohat} it follows $\hat{\rho}(t,x,\lambda) < 1/2\pi$, from which we find $\abs{\hat{\Psi}(0,\hat{x},\lambda)-\hat{\Psi}(0,\hat{y},\lambda)} < \abs{\hat{x}-\hat{y}}/(2\pi)$
. Furthermore, one can explicitly compute $\int\dd{\mu}\varphi(\lambda,\mu) = 2\pi$.

Using these facts we can obtain the following result
\begin{result}
    In the repulsive Lieb-Liniger model the solution to the GHD equation exists and is unique for all times $t$ (see end matter section \ref{sec:precise} for detailed statements). 
\end{result}
To see this, denote the fixed-point map associated to \eqref{equ:fixed_point_hatX} by $C_{t,x}[f] = x+ \int\dd{\mu}\varphi(\lambda,\mu)\hat{\Psi}(0,f(\mu)-v(\mu)t,\mu)$. Now observe that in the supremumsnorm $\norm{f}_\infty = \sup_\lambda \abs{f(\lambda)}$ for any two functions $f_1(\lambda)$ and $f_2(\lambda)$:
\begin{align}
    &\norm{C_{t,x}[f_1]-C_{t,x}[f_2]}_\infty \leq \int\dd{\mu}\varphi(\lambda,\mu)\times\nonumber\\
    &\times |\hat{\Psi}(0,f_1(\mu)-v(\mu)t,\mu)-\hat{\Psi}(0,f_2(\mu)-v(\mu)t,\mu)|\\
    &<\tfrac{1}{2\pi}(\textstyle\int\dd{\mu}\varphi(\lambda,\mu))\norm{f_1(\mu)-f_2(\mu)}_\infty\\
    &= \norm{f_1(\mu)-f_2(\mu)}_\infty.
\end{align}
This is the definition of a contracting fixed-point map. The Banach fixed-point theorem states that a contracting fixed-point map has a unique fixed-point, from which we conclude that $C_{t,x}[f]$ has only one fixed-point $f(\lambda)=\hat{X}(t,x,\lambda)$. Because $\hat{X}(t,x,\lambda)$ is unique, $\Psi(t,x,\lambda)$ is as well and thus $\rho(t,x,\lambda) = \partial_x\Psi(t,x,\lambda)$ is the unique solution to the GHD equation.

Formally this gives us only the existence of bounded functions $\Psi(t,x,\lambda)$, $\hat{X}(t,x,\lambda)$ and thus also $n(t,x,\lambda)$. By solving \eqref{equ:rho_from_n} one can establish that $\rho(t,x,\lambda)$ exists as well as a bounded function. But now note the following: Since $\rho(t,x,\lambda)$ is bounded, its height-field $\Psi(t,x,\lambda)$ is differentiable. Assuming that initially $\hat{\rho}(0,x,\lambda)$ is smooth, we find that $n(t,x,\lambda)$ is differentiable and thus $\rho(t,x,\lambda)$ is also differentiable. This then implies that $\Psi(t,x,\lambda)$ is twice differentiable. By iterating this argument we see that $\rho(t,x,\lambda)$ will be a smooth function. We thus have established the following:

\begin{result}
    In the repulsive Lieb-Liniger model, if the initial state $\rho(0,x,\lambda)$ is smooth, then it will remain smooth for all times. Hence shocks or any other type of gradient singularities are absent (see end matter section \ref{sec:precise} for precise statements).
\end{result}

The Banach fixed-point theorem also includes another, more practical result: It states that the usual fixed-point iteration algorithm (which was for instances used in Fig. \ref{fig:numerics}) converges exponentially fast. 
\begin{result}
    In the repulsive Lieb-Liniger model, the solution to \eqref{equ:fixed_point_Psi} or \eqref{equ:fixed_point_hatX} can be obtained by fixed point iteration from any initial guess $\Psi(t,x,\lambda)$ or $\hat{X}(t,x,\lambda)$ respectively. The iteration algorithm converges exponentially fast.
\end{result}

The arguments can straight-forwardly extended to other models as well. They will also give existence, uniqueness, absence of shock formation and exponentially fast-fixed point convergence if the initial state satisfies $\hat{\rho}(0,\hat{x},\lambda) < C/(\sup_{\lambda}\int\dd{\mu}\abs{\varphi(\lambda,\mu)})$, with $C = 2\pi$ if $\varphi(\lambda)\geq 0$ and $C=\pi$ else (we need to distinguish both cases since for negative $\varphi(\lambda)$ we cannot trivially ensure that $\rho\ind{s}(t,x,\lambda)$ of \eqref{equ:rhos}). This applies to a large family of initial states that are physically relevant. There are also models where this approach cannot be applied, for instance in models with infinitely-many particle species (e.g.~the ``strings" in the XXZ chain \cite{takahashi_1999}), where the total phase-shift (generalized to multiple particle species, see SM~\cite{SM}) diverges $\sum_j\int\dd{\mu}\abs{\varphi_{ij}(\lambda,\mu)} = \infty$. In these cases the fixed-point iteration might diverge. For an example and a workaround strategy to obtain a converging fixed-point iteration see~\cite{quantum_circuit}.

\prlsection{Application: large-scale correlations}
The new fixed-point approach can also be used to understand the dynamics of perturbations. Imagine a situation where the initial state $\rho(0,x,\lambda)$ is perturbed $\rho(0,x,\lambda) \to \rho(0,x,\lambda) + \delta\rho(0,x,\lambda)$. This leads to a perturbation of the fixed-point problems \eqref{equ:fixed_point_Psi} and \eqref{equ:fixed_point_hatX}, and consequently also of its solution. As derived in the SM \cite{SM}, the perturbed solution $\rho(t,x,\lambda) \to \rho(t,x,\lambda) + \delta\rho(t,x,\lambda)$ satisfies
\begin{align}
    \delta \rho(t,x,\lambda) &= \vu{G}_t\delta \rho(0,x,\lambda)\nonumber\\
    &:= \int\dd{y}\dd\mu G_t(x,\lambda|y,\mu)\delta\rho(0,y,\mu)\label{equ:linGHD_kernel_def}
\end{align}
with the evolution kernel
\begin{align}
    G_t(x,\lambda|y,\mu) &= \tfrac{1}{2}\partial_x \Big[\sgn(x-Y(t,y,\lambda)) \times\hspace{1cm}\nonumber\\
    &\times \qty(\delta(\lambda-\mu) - \tfrac{1}{2\pi}n(t,x,\lambda)\varphi(\lambda,\mu))\Big]\upd{drT}.\label{equ:linGHD_kernel}
\end{align}
Here, in the standard terminology of GHD $Y(t,x,\lambda) = \hat{X}^{-1}(t,\hat{X}(0,x,\lambda)-v(\lambda)t)$ is the GHD characteristic, i.e.\ the trajectory of a quasi-particle starting at $x$ with momentum $\lambda$. The operation $f\upd{drT}$ gives the solution to the so called transposed dressing equation
\begin{align}
    f\upd{drT}(\lambda) = f(\lambda) + n(t,x,\lambda)\int\tfrac{\dd{\mu}}{2\pi} \varphi(\lambda,\mu)f\upd{drT}(\mu).
\end{align}

The evolution kernel, which is interpreted as the response of the system to a perturbation $\delta\rho(0,x,\lambda) = \delta(x-y)\delta(\lambda-\mu)$ had been studied previously~\cite[Eq 3.12]{10.21468/SciPostPhys.5.5.054} and computed numerically in \cite{10.21468/SciPostPhysCore.3.2.016}, but \eqref{equ:linGHD_kernel} provides a much simpler and more compact formula. We provide a plot of $G_t(x,\lambda|y,\mu)$ in Fig. \ref{fig:linearized_GHD} and more details about its structure in the end matter section \ref{sec:linGHD}.

\begin{result}
    The dynamics of perturbations of $\rho(t,x,\lambda)$ is given by \eqref{equ:linGHD_kernel_def} and \eqref{equ:linGHD_kernel}.
\end{result}

The evolution of such perturbations are also interesting for the evolution of correlation functions. Following the idea that correlations $\expval{\rho(t,x,\lambda)\rho(s,y,\mu)}\upd{c}$ in the system can be viewed as perturbations spreading on top of the GHD evolution, it is well known that~\cite{10.21468/SciPostPhys.5.5.054}
\begin{align}
    \expval{\rho(t,x,\lambda)\rho(s,y,\mu)}\upd{c} &= \expval{(\vu{G}_t\rho(0,x,\lambda))(\vu{G}_s\rho(0,y,\mu))}.
\end{align}
Here the initial state $\expval{\ldots}$ may or may not be at local equilibrium. From the perspective of hydrodynamics it is interesting to consider an initial local equilibrium state, where $\expval{\rho(0,x,\lambda)\rho(0,y,\mu)}\upd{c} = \delta(x-y)(\text{GGE})$ is uncorrelated for $x\neq y$ (here (GGE) represents the local GGE correlations), and observe whether the correlations change in time. This can be explicitly computed using our results. We find that the evolved equal-time correlations locally look like (see SM \cite{SM}, Eq. \eqrefsm{41} for the explicit formula)
\begin{align}
    \expval{\rho(t,x,\lambda)\rho(t,y,\mu)}\upd{c}\eval_{x\approx y} &= \delta(x-y) (\text{GGE})\nonumber\\
    &+ \sgn(x-y) (\ldots \partial_x \rho(t,x,\lambda))\nonumber\\
    &+ (\mathrm{continuous}).
\end{align}

We indeed find that the GGE correlations are still present as expected. However, there are more terms, so called long range correlations, already observed in~ \cite{PhysRevLett.131.027101}. What is interesting (and a new result) is that these long range correlations jump at $x=y$ proportional to $\partial_x \rho(t,x,\lambda)$. This stems from the sign function in \eqref{equ:linGHD_kernel} \cite{SM} and represents a local \textit{spontaneous} breaking of PT symmetry. Interestingly, this jump develops ``instantaneously'' (from Euler time $t=0^+$)~\footnote{This means the jump develops faster than the hydrodynamic time scale.}. 
\begin{result}
    The equal time two-point correlation function locally is given by a $\delta(x-y)$ peak representing the local GGE plus a finite part which shows a jump proportional to $\partial_x\rho(t,x,\lambda)$ at $x=y$.
\end{result}
This has an influence on diffusive effects, which are discussed in a separate manuscript \cite{diffusion_LR}.

\prlsection{Conclusion} Using ``height fields'', we have constructed new Riemann invariants for the  GHD equation, the universal equation of Euler-scale hydrodynamics in many-body integrable systems. From these, we have obtained new quadratures -- integral equations that solve the GHD equation for fixed position and time. They are not only useful to obtain numerical solutions to the GHD equation, but also provide deeper insights into the dynamics of integrable models: In particular, we have shown that the fixed-point problem Eq.~\eqref{equ:fixed_point_hatX} has a unique solution in the case of the Lieb-Liniger model for all physically relevant states. From this, we have established the main physical properties of the GHD equation, which are in stark contrast to conventional Euler hydrodynamics (the proofs are given in \cite{long}): existence and uniqueness of solutions, and absence of shocks at all times. Establishing such properties in Euler hydrodynamics is in general an extremely difficult task, and this is the first result of this kind for the GHD equation. We believe the techniques developed here may be used to efficiently solve the GHD equation numerically in many more models. Depending on the model, one may need other spaces in which \eqref{equ:fixed_point_hatX} or other quadratures have unique solution; investigating soliton gases \cite{PhysRevE.109.061001} would be particularly useful, as strong mathematical methods are available. It would be interesting to see whether GHD may indeed fail to have unique solutions, such as in models with infinitely-many particle species; and to generalize the results to the presence of external forces \cite{10.21468/SciPostPhys.2.2.014} and diffusive \cite{PhysRevLett.121.160603,10.21468/SciPostPhys.6.4.049,diffusion_LR} and dispersive \cite{DeNardisHigher_2023} corrections. Furthermore, we have shown how to use the new technique to study the evolution of correlation function in integrable systems. The presence of an explicit formula allowed us to establish that even if the system is initially in an local equilibrium state, its correlations quickly differ from the equilibrium correlations. This has an effect on the diffusive correction to GHD, see \cite{diffusion_LR}. One could extend our results to higher-order correlations functions, and use it to simplify the framework of ballistic macroscopic fluctuation theory \cite{10.21468/SciPostPhys.15.4.136} in integrable models. Perhaps Banach-space techniques such as those used here could be employed to study {\em the emergence} of the GHD equation from microscopic dynamics.

{\bf Acknowledgements.}---
FH acknowledges funding from the faculty of Natural, Mathematical \& Engineering Sciences at King's College London. BD was supported by the Engineering and Physical Sciences Research Council (EPSRC) under grant EP/W010194/1. Numerical computations were partially done in Julia~\cite{Julia-2017}, in particular using the amazing ApproxFun.jl library~\cite{ApproxFun.jl-2014}, and ran on the CREATE cluster~\cite{CREATE}.

\bibliographystyle{apsrev4-2}
\bibliography{bib.bib}

\begin{thebibliography}{54}%
\makeatletter
\providecommand \@ifxundefined [1]{%
 \@ifx{#1\undefined}
}%
\providecommand \@ifnum [1]{%
 \ifnum #1\expandafter \@firstoftwo
 \else \expandafter \@secondoftwo
 \fi
}%
\providecommand \@ifx [1]{%
 \ifx #1\expandafter \@firstoftwo
 \else \expandafter \@secondoftwo
 \fi
}%
\providecommand \natexlab [1]{#1}%
\providecommand \enquote  [1]{``#1''}%
\providecommand \bibnamefont  [1]{#1}%
\providecommand \bibfnamefont [1]{#1}%
\providecommand \citenamefont [1]{#1}%
\providecommand \href@noop [0]{\@secondoftwo}%
\providecommand \href [0]{\begingroup \@sanitize@url \@href}%
\providecommand \@href[1]{\@@startlink{#1}\@@href}%
\providecommand \@@href[1]{\endgroup#1\@@endlink}%
\providecommand \@sanitize@url [0]{\catcode `\\12\catcode `\$12\catcode `\&12\catcode `\#12\catcode `\^12\catcode `\_12\catcode `\%12\relax}%
\providecommand \@@startlink[1]{}%
\providecommand \@@endlink[0]{}%
\providecommand \url  [0]{\begingroup\@sanitize@url \@url }%
\providecommand \@url [1]{\endgroup\@href {#1}{\urlprefix }}%
\providecommand \urlprefix  [0]{URL }%
\providecommand \Eprint [0]{\href }%
\providecommand \doibase [0]{https://doi.org/}%
\providecommand \selectlanguage [0]{\@gobble}%
\providecommand \bibinfo  [0]{\@secondoftwo}%
\providecommand \bibfield  [0]{\@secondoftwo}%
\providecommand \translation [1]{[#1]}%
\providecommand \BibitemOpen [0]{}%
\providecommand \bibitemStop [0]{}%
\providecommand \bibitemNoStop [0]{.\EOS\space}%
\providecommand \EOS [0]{\spacefactor3000\relax}%
\providecommand \BibitemShut  [1]{\csname bibitem#1\endcsname}%
\let\auto@bib@innerbib\@empty
\bibitem [{\citenamefont {Castro-Alvaredo}\ \emph {et~al.}(2016)\citenamefont {Castro-Alvaredo}, \citenamefont {Doyon},\ and\ \citenamefont {Yoshimura}}]{PhysRevX.6.041065}%
  \BibitemOpen
  \bibfield  {author} {\bibinfo {author} {\bibfnamefont {O.~A.}\ \bibnamefont {Castro-Alvaredo}}, \bibinfo {author} {\bibfnamefont {B.}~\bibnamefont {Doyon}},\ and\ \bibinfo {author} {\bibfnamefont {T.}~\bibnamefont {Yoshimura}},\ }\href {https://doi.org/10.1103/PhysRevX.6.041065} {\bibfield  {journal} {\bibinfo  {journal} {Phys. Rev. X}\ }\textbf {\bibinfo {volume} {6}},\ \bibinfo {pages} {041065} (\bibinfo {year} {2016})}\BibitemShut {NoStop}%
\bibitem [{\citenamefont {Bertini}\ \emph {et~al.}(2016)\citenamefont {Bertini}, \citenamefont {Collura}, \citenamefont {De~Nardis},\ and\ \citenamefont {Fagotti}}]{bertini2016transport}%
  \BibitemOpen
  \bibfield  {author} {\bibinfo {author} {\bibfnamefont {B.}~\bibnamefont {Bertini}}, \bibinfo {author} {\bibfnamefont {M.}~\bibnamefont {Collura}}, \bibinfo {author} {\bibfnamefont {J.}~\bibnamefont {De~Nardis}},\ and\ \bibinfo {author} {\bibfnamefont {M.}~\bibnamefont {Fagotti}},\ }\href {https://doi.org/10.1103/PhysRevLett.117.207201} {\bibfield  {journal} {\bibinfo  {journal} {Physical review letters}\ }\textbf {\bibinfo {volume} {117}},\ \bibinfo {pages} {207201} (\bibinfo {year} {2016})}\BibitemShut {NoStop}%
\bibitem [{\citenamefont {El}(2003)}]{el2003thermodynamic}%
  \BibitemOpen
  \bibfield  {author} {\bibinfo {author} {\bibfnamefont {G.}~\bibnamefont {El}},\ }\href {https://doi.org/https://doi.org/10.1016/S0375-9601(03)00515-2} {\bibfield  {journal} {\bibinfo  {journal} {Physics Letters A}\ }\textbf {\bibinfo {volume} {311}},\ \bibinfo {pages} {374 } (\bibinfo {year} {2003})}\BibitemShut {NoStop}%
\bibitem [{\citenamefont {Schemmer}\ \emph {et~al.}(2019)\citenamefont {Schemmer}, \citenamefont {Bouchoule}, \citenamefont {Doyon},\ and\ \citenamefont {Dubail}}]{PhysRevLett.122.090601}%
  \BibitemOpen
  \bibfield  {author} {\bibinfo {author} {\bibfnamefont {M.}~\bibnamefont {Schemmer}}, \bibinfo {author} {\bibfnamefont {I.}~\bibnamefont {Bouchoule}}, \bibinfo {author} {\bibfnamefont {B.}~\bibnamefont {Doyon}},\ and\ \bibinfo {author} {\bibfnamefont {J.}~\bibnamefont {Dubail}},\ }\href {https://doi.org/10.1103/PhysRevLett.122.090601} {\bibfield  {journal} {\bibinfo  {journal} {Phys. Rev. Lett.}\ }\textbf {\bibinfo {volume} {122}},\ \bibinfo {pages} {090601} (\bibinfo {year} {2019})}\BibitemShut {NoStop}%
\bibitem [{\citenamefont {Caux}\ \emph {et~al.}(2019)\citenamefont {Caux}, \citenamefont {Doyon}, \citenamefont {Dubail}, \citenamefont {Konik},\ and\ \citenamefont {Yoshimura}}]{10.21468/SciPostPhys.6.6.070}%
  \BibitemOpen
  \bibfield  {author} {\bibinfo {author} {\bibfnamefont {J.-S.}\ \bibnamefont {Caux}}, \bibinfo {author} {\bibfnamefont {B.}~\bibnamefont {Doyon}}, \bibinfo {author} {\bibfnamefont {J.}~\bibnamefont {Dubail}}, \bibinfo {author} {\bibfnamefont {R.}~\bibnamefont {Konik}},\ and\ \bibinfo {author} {\bibfnamefont {T.}~\bibnamefont {Yoshimura}},\ }\href {https://doi.org/10.21468/SciPostPhys.6.6.070} {\bibfield  {journal} {\bibinfo  {journal} {SciPost Phys.}\ }\textbf {\bibinfo {volume} {6}},\ \bibinfo {pages} {070} (\bibinfo {year} {2019})}\BibitemShut {NoStop}%
\bibitem [{\citenamefont {M\o{}ller}\ \emph {et~al.}(2021)\citenamefont {M\o{}ller}, \citenamefont {Li}, \citenamefont {Mazets}, \citenamefont {Stimming}, \citenamefont {Zhou}, \citenamefont {Zhu}, \citenamefont {Chen},\ and\ \citenamefont {Schmiedmayer}}]{PhysRevLett.126.090602}%
  \BibitemOpen
  \bibfield  {author} {\bibinfo {author} {\bibfnamefont {F.}~\bibnamefont {M\o{}ller}}, \bibinfo {author} {\bibfnamefont {C.}~\bibnamefont {Li}}, \bibinfo {author} {\bibfnamefont {I.}~\bibnamefont {Mazets}}, \bibinfo {author} {\bibfnamefont {H.-P.}\ \bibnamefont {Stimming}}, \bibinfo {author} {\bibfnamefont {T.}~\bibnamefont {Zhou}}, \bibinfo {author} {\bibfnamefont {Z.}~\bibnamefont {Zhu}}, \bibinfo {author} {\bibfnamefont {X.}~\bibnamefont {Chen}},\ and\ \bibinfo {author} {\bibfnamefont {J.}~\bibnamefont {Schmiedmayer}},\ }\href {https://doi.org/10.1103/PhysRevLett.126.090602} {\bibfield  {journal} {\bibinfo  {journal} {Phys. Rev. Lett.}\ }\textbf {\bibinfo {volume} {126}},\ \bibinfo {pages} {090602} (\bibinfo {year} {2021})}\BibitemShut {NoStop}%
\bibitem [{\citenamefont {Malvania}\ \emph {et~al.}(2021)\citenamefont {Malvania}, \citenamefont {Zhang}, \citenamefont {Le}, \citenamefont {Dubail}, \citenamefont {Rigol},\ and\ \citenamefont {Weiss}}]{doi:10.1126/science.abf0147}%
  \BibitemOpen
  \bibfield  {author} {\bibinfo {author} {\bibfnamefont {N.}~\bibnamefont {Malvania}}, \bibinfo {author} {\bibfnamefont {Y.}~\bibnamefont {Zhang}}, \bibinfo {author} {\bibfnamefont {Y.}~\bibnamefont {Le}}, \bibinfo {author} {\bibfnamefont {J.}~\bibnamefont {Dubail}}, \bibinfo {author} {\bibfnamefont {M.}~\bibnamefont {Rigol}},\ and\ \bibinfo {author} {\bibfnamefont {D.~S.}\ \bibnamefont {Weiss}},\ }\href {https://doi.org/10.1126/science.abf0147} {\bibfield  {journal} {\bibinfo  {journal} {Science}\ }\textbf {\bibinfo {volume} {373}},\ \bibinfo {pages} {1129} (\bibinfo {year} {2021})},\ \Eprint {https://arxiv.org/abs/https://www.science.org/doi/pdf/10.1126/science.abf0147} {https://www.science.org/doi/pdf/10.1126/science.abf0147} \BibitemShut {NoStop}%
\bibitem [{\citenamefont {Lieb}\ and\ \citenamefont {Liniger}(1963)}]{lieb1963}%
  \BibitemOpen
  \bibfield  {author} {\bibinfo {author} {\bibfnamefont {E.}~\bibnamefont {Lieb}}\ and\ \bibinfo {author} {\bibfnamefont {W.}~\bibnamefont {Liniger}},\ }\href@noop {} {\bibfield  {journal} {\bibinfo  {journal} {Physical Review}\ }\textbf {\bibinfo {volume} {130}},\ \bibinfo {pages} {1605} (\bibinfo {year} {1963})}\BibitemShut {NoStop}%
\bibitem [{\citenamefont {Bouchoule}\ and\ \citenamefont {Dubail}(2022)}]{Bouchoule_2022}%
  \BibitemOpen
  \bibfield  {author} {\bibinfo {author} {\bibfnamefont {I.}~\bibnamefont {Bouchoule}}\ and\ \bibinfo {author} {\bibfnamefont {J.}~\bibnamefont {Dubail}},\ }\href {https://doi.org/10.1088/1742-5468/ac3659} {\bibfield  {journal} {\bibinfo  {journal} {Journal of Statistical Mechanics: Theory and Experiment}\ }\textbf {\bibinfo {volume} {2022}},\ \bibinfo {pages} {014003} (\bibinfo {year} {2022})}\BibitemShut {NoStop}%
\bibitem [{\citenamefont {Suret}\ \emph {et~al.}(2024)\citenamefont {Suret}, \citenamefont {Randoux}, \citenamefont {Gelash}, \citenamefont {Agafontsev}, \citenamefont {Doyon},\ and\ \citenamefont {El}}]{PhysRevE.109.061001}%
  \BibitemOpen
  \bibfield  {author} {\bibinfo {author} {\bibfnamefont {P.}~\bibnamefont {Suret}}, \bibinfo {author} {\bibfnamefont {S.}~\bibnamefont {Randoux}}, \bibinfo {author} {\bibfnamefont {A.}~\bibnamefont {Gelash}}, \bibinfo {author} {\bibfnamefont {D.}~\bibnamefont {Agafontsev}}, \bibinfo {author} {\bibfnamefont {B.}~\bibnamefont {Doyon}},\ and\ \bibinfo {author} {\bibfnamefont {G.}~\bibnamefont {El}},\ }\href {https://doi.org/10.1103/PhysRevE.109.061001} {\bibfield  {journal} {\bibinfo  {journal} {Phys. Rev. E}\ }\textbf {\bibinfo {volume} {109}},\ \bibinfo {pages} {061001} (\bibinfo {year} {2024})}\BibitemShut {NoStop}%
\bibitem [{\citenamefont {Doyon}(2020)}]{10.21468/SciPostPhysLectNotes.18}%
  \BibitemOpen
  \bibfield  {author} {\bibinfo {author} {\bibfnamefont {B.}~\bibnamefont {Doyon}},\ }\href {https://doi.org/10.21468/SciPostPhysLectNotes.18} {\bibfield  {journal} {\bibinfo  {journal} {SciPost Phys. Lect. Notes}\ ,\ \bibinfo {pages} {18}} (\bibinfo {year} {2020})}\BibitemShut {NoStop}%
\bibitem [{\citenamefont {Bastianello}\ \emph {et~al.}(2022)\citenamefont {Bastianello}, \citenamefont {Bertini}, \citenamefont {Doyon},\ and\ \citenamefont {Vasseur}}]{Bastianello_2022}%
  \BibitemOpen
  \bibfield  {author} {\bibinfo {author} {\bibfnamefont {A.}~\bibnamefont {Bastianello}}, \bibinfo {author} {\bibfnamefont {B.}~\bibnamefont {Bertini}}, \bibinfo {author} {\bibfnamefont {B.}~\bibnamefont {Doyon}},\ and\ \bibinfo {author} {\bibfnamefont {R.}~\bibnamefont {Vasseur}},\ }\href {https://doi.org/10.1088/1742-5468/ac3e6a} {\bibfield  {journal} {\bibinfo  {journal} {J. Stat. Mech. Theory Exp.}\ }\textbf {\bibinfo {volume} {2022}},\ \bibinfo {pages} {014001} (\bibinfo {year} {2022})}\BibitemShut {NoStop}%
\bibitem [{\citenamefont {Essler}(2023)}]{ESSLER2023127572}%
  \BibitemOpen
  \bibfield  {author} {\bibinfo {author} {\bibfnamefont {F.~H.}\ \bibnamefont {Essler}},\ }\href {https://doi.org/https://doi.org/10.1016/j.physa.2022.127572} {\bibfield  {journal} {\bibinfo  {journal} {Physica A: Statistical Mechanics and its Applications}\ }\textbf {\bibinfo {volume} {631}},\ \bibinfo {pages} {127572} (\bibinfo {year} {2023})},\ \bibinfo {note} {lecture Notes of the 15th International Summer School of Fundamental Problems in Statistical Physics}\BibitemShut {NoStop}%
\bibitem [{\citenamefont {Spohn}(2024)}]{doi:10.1142/13600}%
  \BibitemOpen
  \bibfield  {author} {\bibinfo {author} {\bibfnamefont {H.}~\bibnamefont {Spohn}},\ }\href {https://doi.org/10.1142/13600} {\emph {\bibinfo {title} {Hydrodynamic Scales of Integrable Many-Body Systems}}}\ (\bibinfo  {publisher} {WORLD SCIENTIFIC},\ \bibinfo {year} {2024})\ \Eprint {https://arxiv.org/abs/https://www.worldscientific.com/doi/pdf/10.1142/13600} {https://www.worldscientific.com/doi/pdf/10.1142/13600} \BibitemShut {NoStop}%
\bibitem [{\citenamefont {Doyon}\ \emph {et~al.}(2017)\citenamefont {Doyon}, \citenamefont {Dubail}, \citenamefont {Konik},\ and\ \citenamefont {Yoshimura}}]{PhysRevLett.119.195301}%
  \BibitemOpen
  \bibfield  {author} {\bibinfo {author} {\bibfnamefont {B.}~\bibnamefont {Doyon}}, \bibinfo {author} {\bibfnamefont {J.}~\bibnamefont {Dubail}}, \bibinfo {author} {\bibfnamefont {R.}~\bibnamefont {Konik}},\ and\ \bibinfo {author} {\bibfnamefont {T.}~\bibnamefont {Yoshimura}},\ }\href {https://doi.org/10.1103/PhysRevLett.119.195301} {\bibfield  {journal} {\bibinfo  {journal} {Phys. Rev. Lett.}\ }\textbf {\bibinfo {volume} {119}},\ \bibinfo {pages} {195301} (\bibinfo {year} {2017})}\BibitemShut {NoStop}%
\bibitem [{\citenamefont {El}\ \emph {et~al.}(2011)\citenamefont {El}, \citenamefont {Kamchatnov}, \citenamefont {Pavlov},\ and\ \citenamefont {Zykov}}]{el2011kinetic}%
  \BibitemOpen
  \bibfield  {author} {\bibinfo {author} {\bibfnamefont {G.}~\bibnamefont {El}}, \bibinfo {author} {\bibfnamefont {A.}~\bibnamefont {Kamchatnov}}, \bibinfo {author} {\bibfnamefont {M.~V.}\ \bibnamefont {Pavlov}},\ and\ \bibinfo {author} {\bibfnamefont {S.}~\bibnamefont {Zykov}},\ }\href@noop {} {\bibfield  {journal} {\bibinfo  {journal} {Journal of Nonlinear Science}\ }\textbf {\bibinfo {volume} {21}},\ \bibinfo {pages} {151} (\bibinfo {year} {2011})}\BibitemShut {NoStop}%
\bibitem [{\citenamefont {Pavlov}\ \emph {et~al.}(2012)\citenamefont {Pavlov}, \citenamefont {Taranov},\ and\ \citenamefont {El}}]{pavlov2012generalized}%
  \BibitemOpen
  \bibfield  {author} {\bibinfo {author} {\bibfnamefont {M.~V.}\ \bibnamefont {Pavlov}}, \bibinfo {author} {\bibfnamefont {V.~B.}\ \bibnamefont {Taranov}},\ and\ \bibinfo {author} {\bibfnamefont {G.~A.}\ \bibnamefont {El}},\ }\href@noop {} {\bibfield  {journal} {\bibinfo  {journal} {Theoretical and Mathematical Physics}\ }\textbf {\bibinfo {volume} {171}},\ \bibinfo {pages} {675} (\bibinfo {year} {2012})}\BibitemShut {NoStop}%
\bibitem [{\citenamefont {Ferapontov}(1991)}]{FERAPONTOV1991112}%
  \BibitemOpen
  \bibfield  {author} {\bibinfo {author} {\bibfnamefont {E.}~\bibnamefont {Ferapontov}},\ }\href {https://doi.org/https://doi.org/10.1016/0375-9601(91)90910-Z} {\bibfield  {journal} {\bibinfo  {journal} {Physics Letters A}\ }\textbf {\bibinfo {volume} {158}},\ \bibinfo {pages} {112} (\bibinfo {year} {1991})}\BibitemShut {NoStop}%
\bibitem [{\citenamefont {Rozdestvenskii}\ and\ \citenamefont {Sidorenko}(1967)}]{lindeg1}%
  \BibitemOpen
  \bibfield  {author} {\bibinfo {author} {\bibfnamefont {B.~L.}\ \bibnamefont {Rozdestvenskii}}\ and\ \bibinfo {author} {\bibfnamefont {A.~D.}\ \bibnamefont {Sidorenko}},\ }\href@noop {} {\bibfield  {journal} {\bibinfo  {journal} {Comput. Math. \& Math. Phys. 7}\ }\textbf {\bibinfo {volume} {7}},\ \bibinfo {pages} {1176–} (\bibinfo {year} {1967})}\BibitemShut {NoStop}%
\bibitem [{\citenamefont {Liu}(1979)}]{lindeg2}%
  \BibitemOpen
  \bibfield  {author} {\bibinfo {author} {\bibfnamefont {T.~P.}\ \bibnamefont {Liu}},\ }\href@noop {} {\bibfield  {journal} {\bibinfo  {journal} {J. Differ. Equ.}\ }\textbf {\bibinfo {volume} {33}},\ \bibinfo {pages} {92–} (\bibinfo {year} {1979})}\BibitemShut {NoStop}%
\bibitem [{\citenamefont {Bressan}(2000)}]{bressan2000hyperbolic}%
  \BibitemOpen
  \bibfield  {author} {\bibinfo {author} {\bibfnamefont {A.}~\bibnamefont {Bressan}},\ }\href {https://global.oup.com/academic/product/hyperbolic-systems-of-conservation-laws-9780198507000?cc=de&lang=en&} {\emph {\bibinfo {title} {Hyperbolic systems of conservation laws: the one-dimensional Cauchy problem}}},\ Vol.~\bibinfo {volume} {20}\ (\bibinfo  {publisher} {Oxford University Press on Demand},\ \bibinfo {year} {2000})\BibitemShut {NoStop}%
\bibitem [{\citenamefont {Doyon}\ \emph {et~al.}(2018{\natexlab{a}})\citenamefont {Doyon}, \citenamefont {Spohn},\ and\ \citenamefont {Yoshimura}}]{DOYON2018570}%
  \BibitemOpen
  \bibfield  {author} {\bibinfo {author} {\bibfnamefont {B.}~\bibnamefont {Doyon}}, \bibinfo {author} {\bibfnamefont {H.}~\bibnamefont {Spohn}},\ and\ \bibinfo {author} {\bibfnamefont {T.}~\bibnamefont {Yoshimura}},\ }\href {https://doi.org/https://doi.org/10.1016/j.nuclphysb.2017.12.002} {\bibfield  {journal} {\bibinfo  {journal} {Nuclear Physics B}\ }\textbf {\bibinfo {volume} {926}},\ \bibinfo {pages} {570} (\bibinfo {year} {2018}{\natexlab{a}})}\BibitemShut {NoStop}%
\bibitem [{\citenamefont {Bulchandani}\ \emph {et~al.}(2017)\citenamefont {Bulchandani}, \citenamefont {Vasseur}, \citenamefont {Karrasch},\ and\ \citenamefont {Moore}}]{PhysRevLett.119.220604}%
  \BibitemOpen
  \bibfield  {author} {\bibinfo {author} {\bibfnamefont {V.~B.}\ \bibnamefont {Bulchandani}}, \bibinfo {author} {\bibfnamefont {R.}~\bibnamefont {Vasseur}}, \bibinfo {author} {\bibfnamefont {C.}~\bibnamefont {Karrasch}},\ and\ \bibinfo {author} {\bibfnamefont {J.~E.}\ \bibnamefont {Moore}},\ }\href {https://doi.org/10.1103/PhysRevLett.119.220604} {\bibfield  {journal} {\bibinfo  {journal} {Phys. Rev. Lett.}\ }\textbf {\bibinfo {volume} {119}},\ \bibinfo {pages} {220604} (\bibinfo {year} {2017})}\BibitemShut {NoStop}%
\bibitem [{\citenamefont {Hübner}\ and\ \citenamefont {Doyon}(2024)}]{long}%
  \BibitemOpen
  \bibfield  {author} {\bibinfo {author} {\bibfnamefont {F.}~\bibnamefont {Hübner}}\ and\ \bibinfo {author} {\bibfnamefont {B.}~\bibnamefont {Doyon}},\ }\href {https://arxiv.org/abs/2411.04922} {\bibinfo {title} {Existence and uniqueness of solutions to the generalized hydrodynamics equation}} (\bibinfo {year} {2024}),\ \Eprint {https://arxiv.org/abs/2411.04922} {arXiv:2411.04922 [math-ph]} \BibitemShut {NoStop}%
\bibitem [{\citenamefont {Doyon}(2018)}]{10.21468/SciPostPhys.5.5.054}%
  \BibitemOpen
  \bibfield  {author} {\bibinfo {author} {\bibfnamefont {B.}~\bibnamefont {Doyon}},\ }\href {https://doi.org/10.21468/SciPostPhys.5.5.054} {\bibfield  {journal} {\bibinfo  {journal} {SciPost Phys.}\ }\textbf {\bibinfo {volume} {5}},\ \bibinfo {pages} {054} (\bibinfo {year} {2018})}\BibitemShut {NoStop}%
\bibitem [{\citenamefont {Møller}\ \emph {et~al.}(2020)\citenamefont {Møller}, \citenamefont {Perfetto}, \citenamefont {Doyon},\ and\ \citenamefont {Schmiedmayer}}]{10.21468/SciPostPhysCore.3.2.016}%
  \BibitemOpen
  \bibfield  {author} {\bibinfo {author} {\bibfnamefont {F.~S.}\ \bibnamefont {Møller}}, \bibinfo {author} {\bibfnamefont {G.}~\bibnamefont {Perfetto}}, \bibinfo {author} {\bibfnamefont {B.}~\bibnamefont {Doyon}},\ and\ \bibinfo {author} {\bibfnamefont {J.}~\bibnamefont {Schmiedmayer}},\ }\href {https://doi.org/10.21468/SciPostPhysCore.3.2.016} {\bibfield  {journal} {\bibinfo  {journal} {SciPost Phys. Core}\ }\textbf {\bibinfo {volume} {3}},\ \bibinfo {pages} {016} (\bibinfo {year} {2020})}\BibitemShut {NoStop}%
\bibitem [{\citenamefont {Doyon}\ \emph {et~al.}(2023{\natexlab{a}})\citenamefont {Doyon}, \citenamefont {Perfetto}, \citenamefont {Sasamoto},\ and\ \citenamefont {Yoshimura}}]{PhysRevLett.131.027101}%
  \BibitemOpen
  \bibfield  {author} {\bibinfo {author} {\bibfnamefont {B.}~\bibnamefont {Doyon}}, \bibinfo {author} {\bibfnamefont {G.}~\bibnamefont {Perfetto}}, \bibinfo {author} {\bibfnamefont {T.}~\bibnamefont {Sasamoto}},\ and\ \bibinfo {author} {\bibfnamefont {T.}~\bibnamefont {Yoshimura}},\ }\href {https://doi.org/10.1103/PhysRevLett.131.027101} {\bibfield  {journal} {\bibinfo  {journal} {Phys. Rev. Lett.}\ }\textbf {\bibinfo {volume} {131}},\ \bibinfo {pages} {027101} (\bibinfo {year} {2023}{\natexlab{a}})}\BibitemShut {NoStop}%
\bibitem [{SM()}]{SM}%
  \BibitemOpen
  \href@noop {} {}\bibinfo {note} {See the Supplemental Material for a discussion of (1) our results with other momentum parametrisations, (2) the equivalence of the various fixed-point problems, (3) our new expressions for correlation functions, and (4) conservation of entropy.}\BibitemShut {Stop}%
\bibitem [{\citenamefont {Doyon}\ \emph {et~al.}(2018{\natexlab{b}})\citenamefont {Doyon}, \citenamefont {Yoshimura},\ and\ \citenamefont {Caux}}]{PhysRevLett.120.045301}%
  \BibitemOpen
  \bibfield  {author} {\bibinfo {author} {\bibfnamefont {B.}~\bibnamefont {Doyon}}, \bibinfo {author} {\bibfnamefont {T.}~\bibnamefont {Yoshimura}},\ and\ \bibinfo {author} {\bibfnamefont {J.-S.}\ \bibnamefont {Caux}},\ }\href {https://doi.org/10.1103/PhysRevLett.120.045301} {\bibfield  {journal} {\bibinfo  {journal} {Phys. Rev. Lett.}\ }\textbf {\bibinfo {volume} {120}},\ \bibinfo {pages} {045301} (\bibinfo {year} {2018}{\natexlab{b}})}\BibitemShut {NoStop}%
\bibitem [{\citenamefont {Doyon}\ \emph {et~al.}(2024)\citenamefont {Doyon}, \citenamefont {H\"ubner},\ and\ \citenamefont {Yoshimura}}]{PhysRevLett.132.251602}%
  \BibitemOpen
  \bibfield  {author} {\bibinfo {author} {\bibfnamefont {B.}~\bibnamefont {Doyon}}, \bibinfo {author} {\bibfnamefont {F.}~\bibnamefont {H\"ubner}},\ and\ \bibinfo {author} {\bibfnamefont {T.}~\bibnamefont {Yoshimura}},\ }\href {https://doi.org/10.1103/PhysRevLett.132.251602} {\bibfield  {journal} {\bibinfo  {journal} {Phys. Rev. Lett.}\ }\textbf {\bibinfo {volume} {132}},\ \bibinfo {pages} {251602} (\bibinfo {year} {2024})}\BibitemShut {NoStop}%
\bibitem [{\citenamefont {Møller}\ and\ \citenamefont {Schmiedmayer}(2020)}]{10.21468/SciPostPhys.8.3.041}%
  \BibitemOpen
  \bibfield  {author} {\bibinfo {author} {\bibfnamefont {F.~S.}\ \bibnamefont {Møller}}\ and\ \bibinfo {author} {\bibfnamefont {J.}~\bibnamefont {Schmiedmayer}},\ }\href {https://doi.org/10.21468/SciPostPhys.8.3.041} {\bibfield  {journal} {\bibinfo  {journal} {SciPost Phys.}\ }\textbf {\bibinfo {volume} {8}},\ \bibinfo {pages} {041} (\bibinfo {year} {2020})}\BibitemShut {NoStop}%
\bibitem [{\citenamefont {Spohn}(1991)}]{Spohn1991}%
  \BibitemOpen
  \bibfield  {author} {\bibinfo {author} {\bibfnamefont {H.}~\bibnamefont {Spohn}},\ }\href {https://doi.org/10.1007/978-3-642-84371-6} {\emph {\bibinfo {title} {Large Scale Dynamics of Interacting Particles}}}\ (\bibinfo  {publisher} {Springer Berlin Heidelberg},\ \bibinfo {year} {1991})\BibitemShut {NoStop}%
\bibitem [{\citenamefont {Pozsgay}\ \emph {et~al.}(2021{\natexlab{a}})\citenamefont {Pozsgay}, \citenamefont {Gombor}, \citenamefont {Hutsalyuk}, \citenamefont {Jiang}, \citenamefont {Pristy\'ak},\ and\ \citenamefont {Vernier}}]{PhysRevE.104.044106}%
  \BibitemOpen
  \bibfield  {author} {\bibinfo {author} {\bibfnamefont {B.}~\bibnamefont {Pozsgay}}, \bibinfo {author} {\bibfnamefont {T.}~\bibnamefont {Gombor}}, \bibinfo {author} {\bibfnamefont {A.}~\bibnamefont {Hutsalyuk}}, \bibinfo {author} {\bibfnamefont {Y.}~\bibnamefont {Jiang}}, \bibinfo {author} {\bibfnamefont {L.}~\bibnamefont {Pristy\'ak}},\ and\ \bibinfo {author} {\bibfnamefont {E.}~\bibnamefont {Vernier}},\ }\href {https://doi.org/10.1103/PhysRevE.104.044106} {\bibfield  {journal} {\bibinfo  {journal} {Phys. Rev. E}\ }\textbf {\bibinfo {volume} {104}},\ \bibinfo {pages} {044106} (\bibinfo {year} {2021}{\natexlab{a}})}\BibitemShut {NoStop}%
\bibitem [{\citenamefont {Pozsgay}\ \emph {et~al.}(2021{\natexlab{b}})\citenamefont {Pozsgay}, \citenamefont {Gombor},\ and\ \citenamefont {Hutsalyuk}}]{PhysRevE.104.064124}%
  \BibitemOpen
  \bibfield  {author} {\bibinfo {author} {\bibfnamefont {B.}~\bibnamefont {Pozsgay}}, \bibinfo {author} {\bibfnamefont {T.}~\bibnamefont {Gombor}},\ and\ \bibinfo {author} {\bibfnamefont {A.}~\bibnamefont {Hutsalyuk}},\ }\href {https://doi.org/10.1103/PhysRevE.104.064124} {\bibfield  {journal} {\bibinfo  {journal} {Phys. Rev. E}\ }\textbf {\bibinfo {volume} {104}},\ \bibinfo {pages} {064124} (\bibinfo {year} {2021}{\natexlab{b}})}\BibitemShut {NoStop}%
\bibitem [{\citenamefont {Doyon}\ \emph {et~al.}(2023{\natexlab{b}})\citenamefont {Doyon}, \citenamefont {Hübner},\ and\ \citenamefont {Yoshimura}}]{doyon2023generalisedtbartdeformationsclassicalfree}%
  \BibitemOpen
  \bibfield  {author} {\bibinfo {author} {\bibfnamefont {B.}~\bibnamefont {Doyon}}, \bibinfo {author} {\bibfnamefont {F.}~\bibnamefont {Hübner}},\ and\ \bibinfo {author} {\bibfnamefont {T.}~\bibnamefont {Yoshimura}},\ }\href {https://arxiv.org/abs/2312.14855} {\bibinfo {title} {Generalised $t\bar{T}$-deformations of classical free particles}} (\bibinfo {year} {2023}{\natexlab{b}}),\ \Eprint {https://arxiv.org/abs/2312.14855} {arXiv:2312.14855 [cond-mat.stat-mech]} \BibitemShut {NoStop}%
\bibitem [{\citenamefont {Croydon}\ and\ \citenamefont {Sasada}(2021)}]{Croydon2021}%
  \BibitemOpen
  \bibfield  {author} {\bibinfo {author} {\bibfnamefont {D.~A.}\ \bibnamefont {Croydon}}\ and\ \bibinfo {author} {\bibfnamefont {M.}~\bibnamefont {Sasada}},\ }\href {https://doi.org/10.1007/s00220-020-03914-x} {\bibfield  {journal} {\bibinfo  {journal} {Communications in Mathematical Physics}\ }\textbf {\bibinfo {volume} {383}},\ \bibinfo {pages} {427} (\bibinfo {year} {2021})}\BibitemShut {NoStop}%
\bibitem [{\citenamefont {Takahashi}(1999)}]{takahashi_1999}%
  \BibitemOpen
  \bibfield  {author} {\bibinfo {author} {\bibfnamefont {M.}~\bibnamefont {Takahashi}},\ }\href {https://doi.org/10.1017/CBO9780511524332} {\emph {\bibinfo {title} {Thermodynamics of One-Dimensional Solvable Models}}}\ (\bibinfo  {publisher} {Cambridge University Press},\ \bibinfo {year} {1999})\BibitemShut {NoStop}%
\bibitem [{\citenamefont {Hübner}\ \emph {et~al.}(2024{\natexlab{a}})\citenamefont {Hübner}, \citenamefont {Vernier},\ and\ \citenamefont {Piroli}}]{quantum_circuit}%
  \BibitemOpen
  \bibfield  {author} {\bibinfo {author} {\bibfnamefont {F.}~\bibnamefont {Hübner}}, \bibinfo {author} {\bibfnamefont {E.}~\bibnamefont {Vernier}},\ and\ \bibinfo {author} {\bibfnamefont {L.}~\bibnamefont {Piroli}},\ }\href {https://arxiv.org/abs/2408.00474} {\bibinfo {title} {Generalized hydrodynamics of integrable quantum circuits}} (\bibinfo {year} {2024}{\natexlab{a}}),\ \Eprint {https://arxiv.org/abs/2408.00474} {arXiv:2408.00474 [cond-mat.stat-mech]} \BibitemShut {NoStop}%
\bibitem [{Note1()}]{Note1}%
  \BibitemOpen
  \bibinfo {note} {This means the jump develops faster than the hydrodynamic time scale.}\BibitemShut {Stop}%
\bibitem [{\citenamefont {Hübner}\ \emph {et~al.}(2024{\natexlab{b}})\citenamefont {Hübner}, \citenamefont {Biagetti}, \citenamefont {Nardis},\ and\ \citenamefont {Doyon}}]{diffusion_LR}%
  \BibitemOpen
  \bibfield  {author} {\bibinfo {author} {\bibfnamefont {F.}~\bibnamefont {Hübner}}, \bibinfo {author} {\bibfnamefont {L.}~\bibnamefont {Biagetti}}, \bibinfo {author} {\bibfnamefont {J.~D.}\ \bibnamefont {Nardis}},\ and\ \bibinfo {author} {\bibfnamefont {B.}~\bibnamefont {Doyon}},\ }\href {https://arxiv.org/abs/2408.04502} {\bibinfo {title} {Diffusive hydrodynamics from long-range correlations}} (\bibinfo {year} {2024}{\natexlab{b}}),\ \Eprint {https://arxiv.org/abs/2408.04502} {arXiv:2408.04502 [cond-mat.stat-mech]} \BibitemShut {NoStop}%
\bibitem [{\citenamefont {Doyon}\ and\ \citenamefont {Yoshimura}(2017)}]{10.21468/SciPostPhys.2.2.014}%
  \BibitemOpen
  \bibfield  {author} {\bibinfo {author} {\bibfnamefont {B.}~\bibnamefont {Doyon}}\ and\ \bibinfo {author} {\bibfnamefont {T.}~\bibnamefont {Yoshimura}},\ }\href {https://doi.org/10.21468/SciPostPhys.2.2.014} {\bibfield  {journal} {\bibinfo  {journal} {SciPost Phys.}\ }\textbf {\bibinfo {volume} {2}},\ \bibinfo {pages} {014} (\bibinfo {year} {2017})}\BibitemShut {NoStop}%
\bibitem [{\citenamefont {{De Nardis}}\ \emph {et~al.}(2018)\citenamefont {{De Nardis}}, \citenamefont {Bernard},\ and\ \citenamefont {Doyon}}]{PhysRevLett.121.160603}%
  \BibitemOpen
  \bibfield  {author} {\bibinfo {author} {\bibfnamefont {J.}~\bibnamefont {{De Nardis}}}, \bibinfo {author} {\bibfnamefont {D.}~\bibnamefont {Bernard}},\ and\ \bibinfo {author} {\bibfnamefont {B.}~\bibnamefont {Doyon}},\ }\href {https://doi.org/10.1103/PhysRevLett.121.160603} {\bibfield  {journal} {\bibinfo  {journal} {Phys. Rev. Lett.}\ }\textbf {\bibinfo {volume} {121}},\ \bibinfo {pages} {160603} (\bibinfo {year} {2018})}\BibitemShut {NoStop}%
\bibitem [{\citenamefont {De~Nardis}\ \emph {et~al.}(2019)\citenamefont {De~Nardis}, \citenamefont {Bernard},\ and\ \citenamefont {Doyon}}]{10.21468/SciPostPhys.6.4.049}%
  \BibitemOpen
  \bibfield  {author} {\bibinfo {author} {\bibfnamefont {J.}~\bibnamefont {De~Nardis}}, \bibinfo {author} {\bibfnamefont {D.}~\bibnamefont {Bernard}},\ and\ \bibinfo {author} {\bibfnamefont {B.}~\bibnamefont {Doyon}},\ }\href {https://doi.org/10.21468/SciPostPhys.6.4.049} {\bibfield  {journal} {\bibinfo  {journal} {SciPost Phys.}\ }\textbf {\bibinfo {volume} {6}},\ \bibinfo {pages} {049} (\bibinfo {year} {2019})}\BibitemShut {NoStop}%
\bibitem [{\citenamefont {De~Nardis}\ and\ \citenamefont {Doyon}(2023)}]{DeNardisHigher_2023}%
  \BibitemOpen
  \bibfield  {author} {\bibinfo {author} {\bibfnamefont {J.}~\bibnamefont {De~Nardis}}\ and\ \bibinfo {author} {\bibfnamefont {B.}~\bibnamefont {Doyon}},\ }\href {https://doi.org/10.1088/1751-8121/acd153} {\bibfield  {journal} {\bibinfo  {journal} {Journal of Physics A: Mathematical and Theoretical}\ }\textbf {\bibinfo {volume} {56}},\ \bibinfo {pages} {245001} (\bibinfo {year} {2023})}\BibitemShut {NoStop}%
\bibitem [{\citenamefont {Doyon}\ \emph {et~al.}(2023{\natexlab{c}})\citenamefont {Doyon}, \citenamefont {Perfetto}, \citenamefont {Sasamoto},\ and\ \citenamefont {Yoshimura}}]{10.21468/SciPostPhys.15.4.136}%
  \BibitemOpen
  \bibfield  {author} {\bibinfo {author} {\bibfnamefont {B.}~\bibnamefont {Doyon}}, \bibinfo {author} {\bibfnamefont {G.}~\bibnamefont {Perfetto}}, \bibinfo {author} {\bibfnamefont {T.}~\bibnamefont {Sasamoto}},\ and\ \bibinfo {author} {\bibfnamefont {T.}~\bibnamefont {Yoshimura}},\ }\href {https://doi.org/10.21468/SciPostPhys.15.4.136} {\bibfield  {journal} {\bibinfo  {journal} {SciPost Phys.}\ }\textbf {\bibinfo {volume} {15}},\ \bibinfo {pages} {136} (\bibinfo {year} {2023}{\natexlab{c}})}\BibitemShut {NoStop}%
\bibitem [{\citenamefont {Bezanson}\ \emph {et~al.}(2017)\citenamefont {Bezanson}, \citenamefont {Edelman}, \citenamefont {Karpinski},\ and\ \citenamefont {Shah}}]{Julia-2017}%
  \BibitemOpen
  \bibfield  {author} {\bibinfo {author} {\bibfnamefont {J.}~\bibnamefont {Bezanson}}, \bibinfo {author} {\bibfnamefont {A.}~\bibnamefont {Edelman}}, \bibinfo {author} {\bibfnamefont {S.}~\bibnamefont {Karpinski}},\ and\ \bibinfo {author} {\bibfnamefont {V.~B.}\ \bibnamefont {Shah}},\ }\href {https://doi.org/10.1137/141000671} {\bibfield  {journal} {\bibinfo  {journal} {SIAM {R}eview}\ }\textbf {\bibinfo {volume} {59}},\ \bibinfo {pages} {65} (\bibinfo {year} {2017})}\BibitemShut {NoStop}%
\bibitem [{\citenamefont {Olver}\ and\ \citenamefont {Townsend}(2014)}]{ApproxFun.jl-2014}%
  \BibitemOpen
  \bibfield  {author} {\bibinfo {author} {\bibfnamefont {S.}~\bibnamefont {Olver}}\ and\ \bibinfo {author} {\bibfnamefont {A.}~\bibnamefont {Townsend}},\ }in\ \href@noop {} {\emph {\bibinfo {booktitle} {Proceedings of the 1st Workshop for High Performance Technical Computing in Dynamic Languages -- HPTCDL `14}}}\ (\bibinfo  {publisher} {{IEEE}},\ \bibinfo {year} {2014})\BibitemShut {NoStop}%
\bibitem [{CRE()}]{CREATE}%
  \BibitemOpen
  \href@noop {} {\bibinfo {title} {{King's College London. (2022). King's Computational Research, Engineering and Technology Environment (CREATE). Retrieved June 30, 2023, from https://doi.org/10.18742/rnvf-m076}}}\BibitemShut {NoStop}%
\bibitem [{\citenamefont {Bonnemain}\ \emph {et~al.}(2022)\citenamefont {Bonnemain}, \citenamefont {Doyon},\ and\ \citenamefont {El}}]{Bonnemain_2022}%
  \BibitemOpen
  \bibfield  {author} {\bibinfo {author} {\bibfnamefont {T.}~\bibnamefont {Bonnemain}}, \bibinfo {author} {\bibfnamefont {B.}~\bibnamefont {Doyon}},\ and\ \bibinfo {author} {\bibfnamefont {G.}~\bibnamefont {El}},\ }\href {https://doi.org/10.1088/1751-8121/ac8253} {\bibfield  {journal} {\bibinfo  {journal} {Journal of Physics A: Mathematical and Theoretical}\ }\textbf {\bibinfo {volume} {55}},\ \bibinfo {pages} {374004} (\bibinfo {year} {2022})}\BibitemShut {NoStop}%
\bibitem [{\citenamefont {Yang}\ and\ \citenamefont {Yang}(1969)}]{yangyang}%
  \BibitemOpen
  \bibfield  {author} {\bibinfo {author} {\bibfnamefont {C.~N.}\ \bibnamefont {Yang}}\ and\ \bibinfo {author} {\bibfnamefont {C.~P.}\ \bibnamefont {Yang}},\ }\href {https://doi.org/10.1063/1.1664947} {\bibfield  {journal} {\bibinfo  {journal} {J. Stat. Phys.}\ }\textbf {\bibinfo {volume} {10}},\ \bibinfo {pages} {1115} (\bibinfo {year} {1969})}\BibitemShut {NoStop}%
\bibitem [{\citenamefont {Zamolodchikov}(1990)}]{Zamolodchikov1990}%
  \BibitemOpen
  \bibfield  {author} {\bibinfo {author} {\bibfnamefont {A.}~\bibnamefont {Zamolodchikov}},\ }\href {https://doi.org/10.1016/0550-3213(90)90333-9} {\bibfield  {journal} {\bibinfo  {journal} {Nucl. Phys. B}\ }\textbf {\bibinfo {volume} {342}},\ \bibinfo {pages} {695} (\bibinfo {year} {1990})}\BibitemShut {NoStop}%
\bibitem [{\citenamefont {Doyon}\ and\ \citenamefont {Spohn}(2017{\natexlab{a}})}]{SciPostPhys.3.6.039}%
  \BibitemOpen
  \bibfield  {author} {\bibinfo {author} {\bibfnamefont {B.}~\bibnamefont {Doyon}}\ and\ \bibinfo {author} {\bibfnamefont {H.}~\bibnamefont {Spohn}},\ }\href {https://doi.org/10.21468/SciPostPhys.3.6.039} {\bibfield  {journal} {\bibinfo  {journal} {SciPost Phys.}\ }\textbf {\bibinfo {volume} {3}},\ \bibinfo {pages} {039} (\bibinfo {year} {2017}{\natexlab{a}})}\BibitemShut {NoStop}%
\bibitem [{\citenamefont {Fendley}\ and\ \citenamefont {Saleur}(1996)}]{PhysRevB.54.10845}%
  \BibitemOpen
  \bibfield  {author} {\bibinfo {author} {\bibfnamefont {P.}~\bibnamefont {Fendley}}\ and\ \bibinfo {author} {\bibfnamefont {H.}~\bibnamefont {Saleur}},\ }\href {https://doi.org/10.1103/PhysRevB.54.10845} {\bibfield  {journal} {\bibinfo  {journal} {Phys. Rev. B}\ }\textbf {\bibinfo {volume} {54}},\ \bibinfo {pages} {10845} (\bibinfo {year} {1996})}\BibitemShut {NoStop}%
\bibitem [{\citenamefont {Doyon}\ and\ \citenamefont {Spohn}(2017{\natexlab{b}})}]{Doyon_2017}%
  \BibitemOpen
  \bibfield  {author} {\bibinfo {author} {\bibfnamefont {B.}~\bibnamefont {Doyon}}\ and\ \bibinfo {author} {\bibfnamefont {H.}~\bibnamefont {Spohn}},\ }\href {https://doi.org/10.1088/1742-5468/aa7abf} {\bibfield  {journal} {\bibinfo  {journal} {Journal of Statistical Mechanics: Theory and Experiment}\ }\textbf {\bibinfo {volume} {2017}},\ \bibinfo {pages} {073210} (\bibinfo {year} {2017}{\natexlab{b}})}\BibitemShut {NoStop}%
\end{thebibliography}%

\clearpage
\onecolumngrid
\section{Other quadrature formulation}\label{sec:other_quadratures}
It is instructive to also derive a related quadrature directly from the GHD equation. The GHD equation for $n(t,x,\lambda)$ is given by~\cite{PhysRevX.6.041065}
\begin{align}
    \partial_t n(t,x,\lambda) + v\upd{eff}(t,x,\lambda)\partial_x n(t,x,\lambda) &= 0\label{equ:GHD_n}.
\end{align}
An important concept is the dressing equation $f\upd{dr}(\lambda)$~\cite{PhysRevX.6.041065}:
\begin{align}
    f\upd{dr}(t,x,\lambda) &= f(\lambda) + \int\tfrac{\dd{\mu}}{2\pi}\varphi(\lambda,\mu)n(t,x,\lambda)f\upd{dr}(t,x,\lambda).
\end{align}
It can be shown that~\cite{PhysRevX.6.041065}
\begin{align}
    2\pi\rho\ind{s}(t,x,\lambda) &= 1\upd{dr}(t,x,\lambda) & v\upd{eff}(t,x,\lambda) = v\upd{dr}(t,x,\lambda)/1\upd{dr}(t,x,\lambda).
\end{align}
and furthermore:
\begin{align}
    \partial_t 1\upd{dr}(t,x,\lambda) + \partial_x v\upd{dr}(t,x,\lambda) &= 0\label{equ:onedr_GHD}.
\end{align}
From this we can follow that there exists a function $K(t,x,\lambda)$ s.t.
\begin{align}
    \partial_x K(t,x,\lambda) &= 1\upd{dr}(t,x,\lambda) &\partial_t K(t,x,\lambda) &= -v\upd{dr}(t,x,\lambda)\label{equ:k_def}.
\end{align}
Using \eqref{equ:k_def} and \eqref{equ:onedr_GHD} in \eqref{equ:GHD_n} we infer $\partial_xK(t,x,\lambda)\partial_tn(t,x,\lambda)=\partial_tK(t,x,\lambda)\partial_xn(t,x,\lambda)$, which is only possible if a function $\tilde{n}(k,\lambda)$ exists s.t.
\begin{align}
    n(t,x,\lambda) = \tilde{n}(K(t,x,\lambda),\lambda)\label{equ:tilden_def}.
\end{align}
Similar to $\hat{\rho}(t,\hat{x},\lambda)$, $\tilde{n}(k,\lambda)$ can be interpreted as expressing the solution to the GHD equation in `spatial coordinate' $K(t,x,\lambda)$ where the GHD equation trivialities $\partial_t \tilde{n}(k,\lambda) = 0$. In fact, if we choose $K(0,0,\lambda) = 0$ we can express $K(t,x,\lambda) = \hat{X}(t,x,\lambda)-v(\lambda)t-\hat{X}(0,0,\lambda)$. Defining $\tilde{N}(k,\lambda) = \int_0^k\dd{k}\tilde{n}(k,\lambda)$ and using \eqref{equ:k_def} we have
\begin{align}
    K(t,x,\lambda) &=\int_{(0,0)}^{(t,x)}1\upd{dr}\dd{x}-v\upd{dr}\dd{t} = x-v(\lambda)t +  \int\tfrac{\dd{\mu}}{2\pi}\varphi(\lambda,\mu)\int_{(0,0)}^{(t,x)}n(t,x,\mu)(1\upd{dr}\dd{x}-v\upd{dr}\dd{t})\nonumber\\ &=x-v(\lambda)t+\int\tfrac{\dd{\mu}}{2\pi}\varphi(\lambda,\mu)\tilde{N}(K(t,x,\mu),\mu) =:G_{t,x}[K].
\end{align}
This is again a fixed-point equation, which can be studied similar to \eqref{equ:fixed_point_hatX}. Once $K(t,x,\lambda)$ has been found the solution $n(t,x,\lambda)$ can be obtained via \eqref{equ:tilden_def}. This formulation has the advantage that it also applies to cases like the partitioning protocoll, where $\rho(t,x,\lambda)$ does not vanish as $x\to -\infty$. Furthermore, we would also like to present the following observation. Under a perturbation of the fixed-point map $G \to G + \delta G$ the fixed point changes as $K \to K + \delta K$ with $\delta K = (\delta G)\upd{dr}$. This means that the the dressing equation is a local version of the fixed-point equation or alternatively the fixed-point equation is a global version of the dressing equation.

\section{Precise statements of existence and uniqueness results}\label{sec:precise}
The following results are proven in a separate paper \cite{long}, but we would like to present them here for completeness.

We assume the following property of the initial state
\begin{align}
    \sup_\lambda \int\dd{\mu}\varphi(\lambda,\mu)\sup_{x} n(0,x,\mu) < C,\label{equ:thm_ass}
\end{align}
where $C=2\pi$ if $\varphi(\lambda,\mu) \geq 0$ or $C=\pi$ else. This is slightly weaker than $n(0,x,\mu)<1/(\sup_{\lambda}\int\tfrac{\dd{\mu}}{2\pi}\varphi(\lambda,\mu))$. In the Lieb-Liniger model it allows us to include all physically relevant states studied until now, zero-entropy states \cite{PhysRevLett.119.195301}; in particular, $n(0,x,\mu)$ can reach its maximal value $1$.

Note that we also allow initial states with discontinuities (this is for instance relevant for the important case of the partitioning protocol). In this case the solution $\rho(t,x,\lambda)$ will not be differentiable and thus \eqref{equ:GHD_conservation} does not make sense. It only makes sense in a weak form; for this purpose, we use the global form of the conservation law:

\begin{equation}\label{equ:weak}\begin{aligned}
        \int_{x_1}^{x_2}\dd x\,\big(\rho(t_2,x,\lambda) - \rho(t_1,x,\lambda)\big)&=
        \int_{t_1}^{t_2}\dd t\,\big(v^{\rm eff}(t,x_1,\lambda)\rho(t,x_1,\lambda) - v^{\rm eff}(t,x_2,\lambda)\rho(t,x_2,\lambda)\big).
    \end{aligned}
\end{equation}

Then we obtain the following statements:

\begin{enumerate}
	\item Existence and uniqueness of weak solutions: If \eqref{equ:thm_ass} and $\sup_{x,\lambda} |v(\lambda) n(0,x,\lambda)|<\infty$, then the solution to \eqref{equ:weak} exists and is unique, and satisfies $\sup_{t,x,\lambda} n(t,x,\lambda)= \sup_{x,\lambda} n(0,x,\lambda)$.
	\item Absence of shocks: If further $n(0,x,\lambda)$ is smooth in $x$ and $\sup_{x,\lambda}(1+|\lambda|^{r+1})|\partial_x^r n(0,x,\lambda)|<\infty$ for every $r$, then the (strong) solution to \eqref{equ:GHD_conservation} exists, is unique, and is smooth as a function of $(t,x)$.
\end{enumerate}

As part of this, we have also shown that the solution $v\upd{eff}(t,x,\lambda)$ to \eqref{equ:veff} exists and is unique for all $x,t$, and that the dressing operation is well defined.

\section{Further details on the evolution of perturbations}\label{sec:linGHD}
\begin{figure}[!h]
    \centering
    \includegraphics{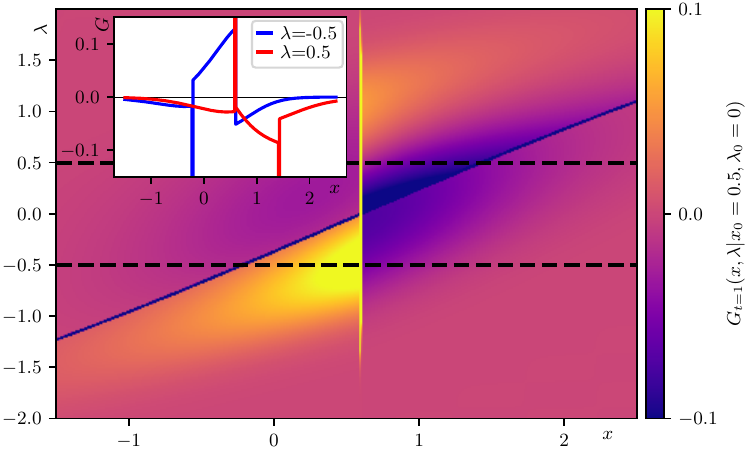}
    \caption{Solution kernel $G_{t=1}(x,\lambda|x_0=0.5,\lambda_0=0)$ to the linearized GHD equation of the (repulsive) Lieb-Liniger model $c=1$ starting from a delta peak at $x_0=1,\lambda_0=0$ at $t=0$. The background density at initial time is given by $\rho(t=0,x,\lambda) = e^{-x^2/2-\lambda^2/2}/4$. The inset shows cuts at $\lambda = \pm 1/2$ (indicated by dashed lines). The solution kernel has two delta functions in $x$, a positive one at the position of the GHD characteristic starting at $(x_0,\lambda_0)$, which is independent of $\lambda$ and a negative one at the position of the GHD characteristic starting at $(x_0,\lambda)$. At the locations of the $\delta$ peaks the non-singular part of $G(t=1,x,\lambda|x_0=0.5,\lambda_0=0)$ has a jump given by \eqref{equ:linGHD_kernel_sing}.}
    \label{fig:linearized_GHD}
\end{figure}
It is instructive to extract the singular part from the evolution kernel $G(t,x,\lambda|y,\mu)$:
\begin{align}\label{equ:linGHD_kernel_sing}
	G(t,x,\lambda|y,\mu) &= \delta(x-Y(t,y,\mu))\qty[\delta(\lambda-\mu)]\upd{drT} -\delta(x-Y(t,y,\lambda))\tfrac{1}{2\pi}n(t,x,\lambda)\varphi(\lambda,\mu)\nonumber\\
    &-\tfrac{1}{2} \sgn(x-Y(t,y,\lambda))\tfrac{1}{2\pi}\partial_x n(t,x,\lambda)\varphi(\lambda,\mu)) +\tfrac{1}{2}\sgn(x-Y(t,y,\mu))\qty[\partial_xn(t,x,\lambda)\vu{T}\qty[\delta(\lambda-\mu)]\upd{drT}]\upd{drT}\nonumber\\
    &+\mathrm{(continuous)}.
\end{align}
As we can read off, the solution kernel consists of two $\delta$ functions with different coefficients that have opposite signs. The positive one is independent of $\lambda$ and located at the GHD characteristic $Y(t,y,\mu)$ corresponding to $\mu$. In addition there is also a negative $\delta$ peak at the characteristic $Y(t,y,\lambda)$ of a particle with rapidity $\lambda$ starting at $y$. In addition there is a background density, which jumps at the location of the $\delta$ peaks (also see Fig. \ref{fig:linearized_GHD} where we give a plot of the solution kernel for a specific example). The jump of this non-singular part is precisely what causes the jump in the evolution of the correlation functions as well.

\def\eqrefmt#1{(#1MT)}

\onecolumngrid
\newpage 
\appendix

\begin{center}
{\Large Supplemental material for ``A new quadrature for the generalized hydrodynamics equation and absence of shocks in the Lieb-Liniger model''
}
\end{center}

\setcounter{equation}{0}
\setcounter{figure}{0}
\setcounter{table}{0}
\setcounter{page}{1}
\renewcommand{\theequation}{S\arabic{equation}}
\setcounter{figure}{0}
\renewcommand{\thefigure}{S\arabic{figure}}
\renewcommand{\thepage}{S\arabic{page}}
\renewcommand{\thesection}{S\arabic{section}}
\renewcommand{\thetable}{S\arabic{table}}
\makeatletter

\renewcommand{\thesection}{\arabic{section}}
\renewcommand{\thesubsection}{\thesection.\arabic{subsection}}
\renewcommand{\thesubsubsection}{\thesubsection.\arabic{subsubsection}}

We refer to equations from the main text as \eqrefmt{a} and to equations within the SM as (a).
\section{Fixed point equations with arbitrary momentum function}
In many models there are quasi-particles of more than one type. This means that the state is described by individual densities $\rho_i(t,x,\lambda)$ of particle types $i=1,2,\ldots$. Moreover the momentum of a quasi-particle $p$ is not necessarily equal to its rapidity $\lambda$, but is instead given by a function $p=P_i(\lambda)$. The formulas given in the main text assume that $P_i(\lambda)=\lambda$. This is not a problem, since GHD can always be brought into a form where $P_i(\lambda)=\lambda$ (GHD is co-variant under a change of parametrisation of the momentum, hence $P_i(\lambda)$ is a gauge freedom of GHD~\cite{10.21468/SciPostPhys.5.5.054,Bonnemain_2022}). In general, however, under change of parametrisation, $\varphi_{ij}(\lambda,\mu)$ is modified \cite{10.21468/SciPostPhys.5.5.054} and might take a simpler form in some parametrization and thus it is useful to keep $P_i(\lambda)$ explicit in some models. Here we note the fixed-point equations in the case of general momentum function.

In general, for quasi-particles with momentum $P_i(\lambda)$ and energy $E_i(\lambda)$ and scattering shifts $\varphi_{ij}(\lambda,\mu)$, the GHD equation reads
\begin{align}
    \partial_t \rho_i(t,x,\lambda) + \partial_x(v\upd{eff}_i(t,x,\lambda)\rho_i(t,x,\lambda)) &= 0
\end{align}
where the effective velocity satisfies
\begin{align}
    v\upd{eff}_i(t,x,\lambda) = \frac{ E_i'(\lambda)}{P_i'(\lambda)} + \sum_j\int\dd{\mu} \frac{\varphi_{ij}(\lambda,\mu)}{P_i'(\lambda)}\rho_{j}(t,x,\mu)(v\upd{eff}_j(t,x,\mu)-v\upd{eff}_i(t,x,\lambda)),
\end{align}
with $f'(\lambda)=\dv{\lambda}f(\lambda)$.

One defines the `density of available space' as $\rho_{{\rm s},i}(t,x,\lambda)
= {P'}_i(\lambda)/(2\pi)+(\vu{T}\rho)_i (t,x,\lambda)$ and the occupation function as $n_i(\lambda) = \rho_i(\lambda)/\rho_{{\rm s},i}(\lambda)$. One can show $\rho_{{\rm s},i}(\lambda) = {P'}\upd{dr}_i(\lambda)/(2\pi)$ and $v\upd{eff}_i(\lambda) = {E'}\upd{dr}_i(\lambda)/{P'}\upd{dr}_i(\lambda)$.

In analogy to the case $P_i(\lambda)=\lambda$ we can define the normal coordinates
\begin{align}
    \hat{X}_i(t,x,\lambda)= {P'}_i(\lambda)x + \int_{-\infty}^x\dd{y}\sum_j\int\dd{\mu}\varphi_{ij}(\lambda,\mu)\rho_{j}(t,y,\mu),
\end{align}
in which $\hat{\rho}_{i}(t,\hat{X}_i(t,x,\lambda),\lambda)\dd{\hat x}= \rho_{i}(t,x,\lambda)\dd{x}$ satisfies $\partial_t \hat{\rho}_{i}(t,\hat{x},\lambda) + E'_i(\lambda)\partial_{\hat{x}}\hat{\rho}_{i}(t,\hat{x},\lambda) = 0$. Now $\Psi_i(t,x,\lambda) = \int_{-\infty}^x\dd{x'}\rho_i(t,x,\lambda)$ satisfies the following fixed-point equation:
\begin{align}
	& \Psi_i(t,x,\lambda) = \hat \Psi_i\Big(0,P'_i(\lambda)x-E'_i(\lambda)t + \sum_j \int \dd{\mu}\varphi_{ij}(\lambda,\mu)\Psi_j(t,x,\mu)\,,\,\lambda\Big),
\end{align}
where $\hat \hat{\Psi}_i(0,\hat{x},\lambda)$ can again be determined from the initial condition.

Alternatively, we can define $K_i(t,x,\lambda)$ by $\partial_x K_i(t,x,\lambda)={P'}\upd{dr}_i(t,x,\lambda)$ and $\partial_t K_i(t,x,\lambda)=-{E'}\upd{dr}_i(t,x,\lambda)$. Then the solution is given by $n_i(t,x,\lambda) = \tilde{n}_i(K_i(t,x,\lambda),\lambda)$, where $\hat{n}_i(k,\lambda)$ is again fixed by the initial condition. The $K_i(t,x,\lambda)$ satisfies the following fixed-point equation:
\begin{align}
    K_i(t,x,\lambda) = P'_i(\lambda)x - E'_i(\lambda)t + \sum_j\int\tfrac{\dd{\mu}}{2\pi}\varphi_{ij}(\lambda,\mu) \tilde{N}_i(K_i(t,x,\mu),\mu) =: G_{t,x}[K(t,x)],
\end{align}
where $\tilde{N}_i(k,\mu) = \int_0^k\dd{k}\tilde{n}_i(k,\lambda)$. 

\section{Application to the partitioning protocol}
One of the signature problems of GHD is the partitioning protocol: Two states $\rho\ind{\pm}(\lambda)$ are joined at $x=0$, s.t. $\rho(x,\lambda) = \rho_{\sgn(x)}(\lambda)$. From these we can compute $n_{\pm}(\lambda)$ and $1\upd{dr}_{\pm}(\lambda)$. We can write:
\begin{align}
    K(0,x,\lambda) &= 1\upd{dr}_{\sgn(x)}(\lambda)x &
    \hat{n}(k,\lambda) &= n_{\sgn(k)}(\lambda).
\end{align}
We can write the fixed point equation for $K$ \eqrefmt{28} as:
\begin{align}
    K(t,x,\lambda) &= x-v(\lambda) t + \int\tfrac{\dd{\mu}}{2\pi}\varphi(\lambda,\mu)\qty[n_{\sgn(K(t,x,\mu))}(\mu)K(t,x,\mu)].
\end{align}
Note that by defining $\zeta = x/t$ and $\tilde{K}(\zeta,\lambda)=K(t,x,\lambda)/t$ we have for all $t>0$:
\begin{align}
    \tilde{K}(\zeta,\lambda) &= \zeta-v(\lambda) + \int\tfrac{\dd{\mu}}{2\pi}\varphi(\lambda,\mu)\qty[n_{\sgn(\tilde{K}(t,x,\mu))}(\mu)\tilde{K}(\zeta,\mu)].
\end{align}
Therefore the solution only depends on the ray $\zeta$, which is a standard property of the partitioning protocol~\cite{bressan2000hyperbolic,PhysRevX.6.041065,10.21468/SciPostPhysLectNotes.18}. Furthermore, note that the solution:
\begin{align}
    n(t,x,\lambda) = \tilde{n}(\zeta,\lambda) = n_{\sgn(\tilde{K}(\zeta,\lambda))}(\lambda),
\end{align}
depends only on the sign of $\tilde{K}(\zeta,\lambda)$. Since $K(t,x,\lambda)$ is constant along a GHD characteristic, the occupation function coincides with the left (right) initial state if $K(t,x,\lambda)$ ($K(t,x,\lambda)>0$) and the jump $K(t,x,\lambda)=0$ propagates along a GHD characteristic. This is again a confirmation of a known result~\cite{PhysRevX.6.041065,10.21468/SciPostPhysLectNotes.18}.

\section{Further Riemann-invariants of the GHD equation}

As noted in \cite{10.21468/SciPostPhys.6.6.070}, $\rho^f(t,x,\lambda) = f(n(t,x,\lambda),\lambda)\rho_{{\rm s}}(t,x,\lambda)$ is a conserved density satisfying \eqref{equ:GHD_conservation} for any function $f(n,\lambda)$, because $\rho(t,x,\lambda)=\rho_{{\rm s}}(t,x,\lambda)n(t,x,\lambda)$ satisfies the conservation law \eqrefmt{1}, and $n(t,x,\lambda)$ the transport law \eqrefmt{22}, {\em with the same effective velocity}. The fact that the same effective velocity appears ``within and without" the spatial derivative is not expected in generic hydrodynamic equations, even if Riemann invariants are known; but it holds in GHD. But then, this implies that any {\em height field} $\Psi^f(t,x,\lambda) = \int_{-\infty}^x \dd{x}' \rho^f(t,x',\lambda)$ is a Riemann invariant,
\begin{equation}
	\partial_t \Psi^f(t,x,\lambda) + v^{\rm eff}(t,x,\lambda)\partial_x \Psi^f(t,x,\lambda) = 0.
\end{equation}
A height field is a ``potential'' for the continuity equation: as $\partial_t \rho^f + \partial_x (v^{\rm eff} \rho^f) = 0$, there exists $\Psi^f$ such that $\partial_x \Psi^f = \rho^f$ and $\partial_t \Psi^f = -v^{\rm eff}\rho^f$. Using path-independence, $\Psi^f = \int_{(t_0,x_0)}^{(t,x)} \big(\rho^f \dd x - v^{\rm eff}\rho^f \dd t\big)$; here we have chosen $x_0=-\infty$ assuming asymptotic vanishing.

Therefore, under the GHD evolution, $\Psi^f(t,x,\lambda) = \Psi^f(0,u_i(t,x,\lambda),\lambda)$. This gives infinitely-many new quadratures. As an example, with $f(n)=n$ one obtains $\Psi(t,x,\lambda)$ as introduced in the main text. 

One can even go beyond this. Using the new Riemann invariants one can easily show that 
\begin{align}
    \int\dd{x} F(n(t,x,\lambda),\Psi^f(t,x,\lambda),\lambda)\rho_{{\rm s}}(t,x,\lambda)
\end{align}
is as well a conserved quantity. This way one can recursively construct new Riemann invariants by considering height fields of these conserved densities, use them to find new Riemann invariants, etc.

\section{Equivalence of both fixed-point problems}
The two fixed point problems are related as follows. Note that $\dv{x}\hat{X}(t,x,\lambda) = 1\upd{dr}(t,x,\lambda) = \dv{x}K(t,x,\lambda)$, thus they only differ by a constant:
\begin{align}
    \hat{X}(t,x,\lambda) &= K(t,x,\lambda) + C(t,\lambda).
\end{align}
We can explicitly compute:
\begin{align}
    \dv{t} (\hat{X}(t,x,\lambda)-K(t,x,\lambda)) &= -2\pi\vu{T}v\upd{eff}(t,x,\lambda)\rho(t,x,\lambda) + v\upd{dr}(t,x,\lambda) = -\vu{T}n(t,x,\lambda)v\upd{dr}(t,x,\lambda) + v\upd{dr}(t,x,\lambda) = v(\lambda)
\end{align}
from which it follows:
\begin{align}
    \hat{X}(t,x,\lambda) &= K(t,x,\lambda) + v(\lambda) t + C(\lambda),
\end{align}
where we fixed the additional constant, s.t. $K(0,0,\lambda) = 0$.

From \eqrefmt{11} we find:
\begin{align}
    \hat{X}(t,x,\lambda) &= x+\vu{T}\hat{N}(\hat{X}(t,x,\lambda)-v_i(\lambda)t,\lambda).
\end{align}
Thus,
\begin{align}
    K(t,x,\lambda) &= x-v(\lambda) t - C(\lambda) + \vu{T}\hat{N}(K(t,x,\lambda) + C(\lambda),\lambda). 
\end{align}
This has, up to the constant of integration $C(\lambda)$, which can by fixed by demanding $K(t,x,\lambda)=0$, the fixed point equation for $K$. 

\section{Conservation of charge and entropy}
The GHD equation has the fundamental property that it conserves the following quantities:
\begin{align}
	S(t) &= \int\dd{x}\dd{\lambda} \rho_{{\rm s}}(t,x,\lambda) \alpha(n(t,x,\lambda),\lambda),
\end{align}
where $\alpha(n,\lambda)$ is an arbitrary function. This follows immediately from the GHD equation~\cite{10.21468/SciPostPhys.6.6.070}, but we can also explicitly show it using the obtained solution: Recall that $n(t,x,\lambda) = \tilde{n}(K(t,x,\lambda),\lambda)$ and $\partial_x K(t,x,\lambda) = 1\upd{dr}(t,x,\lambda) = 2\pi \rho_{{\rm s}}(t,x,\lambda)$ and observe:
\begin{align}
	S(t) &= \tfrac{1}{2\pi}\int\dd{x}\dd{\lambda} \partial_x K(t,x,\lambda) \alpha(\tilde{n}(K(t,x,\lambda),\lambda),\lambda) = \tfrac{1}{2\pi}\int\dd{k}\dd{\lambda} \alpha(\tilde{n}(k,\lambda),\lambda), 
\end{align}
which is independent of time.

We would like to discuss two important special cases of $\alpha(n,\lambda)$: First, in case $\alpha(n,\lambda) = f(\lambda)$, $S(t)$ is a microscopically conserved charge of the integrable model. For instance, in the Lieb-Liniger model $f(\lambda) = 1$,$f(\lambda) = \lambda$ and $f(\lambda) = \lambda^2$ are the total particle number, momentum and energy respectively.

The other quantities are entropies: From the general theory of integrability we know that the entropy density is given by the thermodynamic Bethe ansatz \cite{yangyang,Zamolodchikov1990,takahashi_1999} (a discussion of its most general form is found in \cite{10.21468/SciPostPhys.6.4.049}):
\begin{align}
	s = -\int\dd{\lambda}\rho_{{\rm s}}(\lambda) \gamma(n(\lambda)).
\end{align}
Here $\gamma(n)$ is a function that depends on the statistics of particles. For instance classical particles we have $\gamma(n) = n\log n -n$, while for quantum particles with fermionic statistics its $\gamma(n) = n\log n + (1-n)\log(1-n)$. We can see that if we set $\alpha(n,\lambda)=-\gamma(n)$, we obtain the conservation of entropy. Note that the GHD equation conserves the entropy corresponding to any particle type. This is because it is possible to obtain the same GHD equation in, say, a quantum and a classical model. Therefore the GHD equation has to be agnostic to the particle statistics and has to conserve all possible entropies $\gamma(n)$.

\section{Solution to the linearized Euler equation}

\subsection{The kernel from linear response}

Assume that we change the initial condition to the GHD equation by some small perturbation $\rho_(x,\lambda) \to \rho(0,x,\lambda) + \delta \rho(0,x,\lambda)$. Using the fixed point equations we can find the correction to the solution. First, the perturbation $\delta \rho(0,x,\lambda)$ introduces a perturbation of the height field and the coordinate transformation at initial times:
\begin{align}
    \delta \Psi(0,x,\lambda) &= \int_{-\infty}^x\dd{y}\delta \rho(0,y,\lambda)\\
    \delta \hat{X}(0,x,\lambda) &=  \int\dd{\mu}\varphi(\lambda,\mu)\delta \Psi(0,x,\mu).
\end{align}
Next, let us use $\hat{\Psi}(0,\hat{X}(0,x,\lambda),\lambda) = \Psi_i(0,x,\lambda)$ to obtain its perturbation:
\begin{align}
    \delta \hat{\Psi}(0,\hat{x},\lambda) &= \delta \Psi(0,X(0,\hat{x},\lambda),\lambda) - \hat{\rho}(0,\hat{x},\lambda) \int\dd{\mu}\varphi(\lambda,\mu)\delta \Psi(0,X(0,\hat{x},\lambda),\mu)\\
    &= \int\dd{\mu}\qty[\delta(\lambda-\mu)-n(0,x,\lambda)\tfrac{\varphi(\lambda,\mu)}{2\pi}]\delta \Psi(0,x,\mu)\eval_{x=X(0,\hat{x},\lambda)}\\
    &= \qty[(\vb{1}-n(0,x,\lambda)\vu{T}) \delta \Psi(0,x,\lambda)]_{x=X(0,\hat{x},\lambda)}.
\end{align}
Here $X(t,\hat{x},\lambda)$ is the inverse function of $\hat{X}(t,x,\lambda)$ in $x$ and we defined the scattering kernel
\begin{align}
    \vu{T}f(\lambda) = \int\tfrac{\dd{\mu}}{2\pi}\varphi(\lambda,\mu)f(\mu), 
\end{align}
and we are going to use the convention that operators (for instance $\vu{T}$ and $\cdot\upd{dr}$) act on $\lambda$. Similarly we obtain at later time $t$:
\begin{align}
    \delta \Psi(t,x,\lambda) &= \delta \hat{\Psi}(t,\hat{X}(t,x,\lambda),\lambda) + n(t,x,\lambda) \vu{T}\delta \Psi(t,x,\lambda),
\end{align}
which can be formally solved as:
\begin{align}
    \delta \Psi(t,x,\lambda) &= (\vb{1}-n(t,x,\lambda)\vu{T})^{-1} \delta \hat{\Psi}(t,\hat{X}(t,x,\lambda),\lambda) = \qty[\delta \hat{\Psi}(t,\hat{X}(t,x,\lambda),\lambda)]\upd{drT}.
\end{align}
Here we identified $\cdot\upd{drT} = (\vb{1}-n(t,x,\lambda)\vu{T})^{-1}$.

Since the evolution in $\hat{x}$ coordinates is trivial, the height field at late times is given by:
\begin{align}
    \delta \Psi(t,x,\lambda) &= (\vb{1}-n(t,x,\lambda)\vu{T})^{-1} \qty(\qty[(\vb{1}-n(0,y,\lambda)\vu{T})\delta \Psi(0,y,\lambda)]_{y = u(t,x,\lambda)}),
\end{align}
where 
\begin{align}
    u(t,x,\lambda) = X(0,\hat{X}(t,x,\lambda)-v(\lambda)t,\lambda)\label{equ:sm_linGHD_Y0}
\end{align}
is the starting point (at time $t=0$) of the GHD characteristic ending at $x$ at time $t$ with rapidity $\lambda$. By taking a derivative we can compute $\delta \rho(t,x,\lambda) = \partial_x\delta \Psi(t,x,\lambda)$.

If the initial state is given by $\delta \rho(0,x,\lambda) = \delta(x-x_0)\delta(\lambda-\lambda_0)$ the solution at later times is therefore given by
\begin{align}
    \delta \rho(t,x,\lambda)=G(t,x,\lambda|x_0,\lambda_0)&= \partial_x \qty[\theta(u(t,x,\lambda)-x_0)\qty(\delta(\lambda-\lambda_0) - \tfrac{1}{2\pi}n(t,x,\lambda)\varphi(\lambda,\lambda_0))]\upd{drT}.
\end{align}
We can decompose $\theta(x) = \tfrac{1}{2}+\tfrac{1}{2}\sgn{x}$ and find:
\begin{align}
    G(t,x,\lambda|x_0,\lambda_0)=G(t,x,\lambda|x_0,\lambda_0)&= \tfrac{1}{2}\partial_x \qty[\sgn(u(t,x,\lambda)-x_0)\qty(\delta(\lambda-\lambda_0) - \tfrac{1}{2\pi}n(t,x,\lambda)\varphi(\lambda,\lambda_0))]\upd{drT}.
\end{align}
Note that, since $u(t,x,\lambda)$ is strictly increasing, we can write:
\begin{align}\label{equ:linGHD_kernel}
    G(t,x,\lambda|x_0,\lambda_0) &= \tfrac{1}{2}\partial_x \qty[\sgn(x-Y(t,x_0,\lambda))\qty(\delta(\lambda-\lambda_0) - \tfrac{1}{2\pi}n(t,x,\lambda)\varphi(\lambda,\lambda_0))]\upd{drT}.
\end{align}
Here $Y(t,x_0,\lambda)$ is the position of a GHD characteristic starting from $x_0$ with $\lambda$, i.e.\ the solution to the ODE
\begin{align}
    \dv{t} Y(t,x_0,\lambda) &= v\upd{eff}(t,Y(t,x_0,\lambda),\lambda) & Y(t=0,x_0,\lambda)&=x_0.
\end{align}
As a next step let us take the derivative inside the transpose dressing:
\begin{align}
    G(t,x,\lambda|x_0,\lambda_0) &= \qty[\delta(x-Y(t,x_0,\lambda))\qty(\delta(\lambda-\lambda_0) - \tfrac{1}{2\pi}n(t,x,\lambda)\varphi(\lambda,\lambda_0))]\upd{drT}\nonumber\\
    &-\tfrac{1}{2} \qty[\sgn(x-Y(t,x_0,\lambda))\tfrac{1}{2\pi}\partial_x n(t,x,\lambda)\varphi(\lambda,\lambda_0))]\upd{drT}\nonumber\\
    &+\tfrac{1}{2}\qty[\partial_xn(t,x,\lambda)\vu{T}\qty[\sgn(x-Y(t,x_0,\lambda))\qty(\delta(\lambda-\lambda_0) - \tfrac{1}{2\pi}n(t,x,\lambda)\varphi(\lambda,\lambda_0))]\upd{drT}]\upd{drT}
\end{align}

In order to understand the main features of this state, we now expand $f\upd{drT} = \sum_{a=0}^\infty \qty(n(t,x,\lambda)\vu{T})^a f$ and only keep non-continuous terms:
\begin{align}\label{equ:linGHD_kernel_sing}
    G(t,x,\lambda|x_0,\lambda_0) &= \delta(x-Y(t,x_0,\lambda_0))\qty[\delta(\lambda-\lambda_0)]\upd{drT}\nonumber\\
    &-\delta(x-Y(t,x_0,\lambda))\tfrac{1}{2\pi}n(t,x,\lambda)\varphi(\lambda,\lambda_0)\nonumber\\
    &-\tfrac{1}{2} \sgn(x-Y(t,x_0,\lambda))\tfrac{1}{2\pi}\partial_x n(t,x,\lambda)\varphi(\lambda,\lambda_0))\nonumber\\
    &+\tfrac{1}{2}\sgn(x-Y(t,x_0,\lambda_0))\qty[\partial_xn(t,x,\lambda)\vu{T}\qty[\delta(\lambda-\lambda_0)]\upd{drT}]\upd{drT} +\mathrm{(continuous)}.
\end{align}

\subsection{Evolution of  correlation functions}
As explained in the main text the (Euler scale) correlations of charge densities $\expval{\rho(t,x,\lambda)\rho(s,y,\mu)}\upd{c}$ evolve according to the linearized Euler equation in both components. Using the solution kernel \eqref{equ:linGHD_kernel} we can thus write:
\begin{align}
    \expval{\rho(t,x,\lambda)\rho(s,y,\mu)}\upd{c} &= \int\dd{x_0}\dd{\lambda_0}\dd{y_0}\dd{\mu_0} G(t,x,\lambda|x_0,\lambda_0)G(s,y,\mu|y_0,\mu_0)\expval{\rho(0,x_0,\lambda_0)\rho(0,y_0,\mu_0)}\upd{c}.
\end{align}

A particularly interesting case is the equal time correlation function $t=s$:
\begin{align}\label{equ:corr_equal_time}
    \expval{\rho(t,x,\lambda)\rho(t,y,\mu)}\upd{c} &= \sum\int\dd{x_0}\dd{\lambda_0}\dd{y_0}\dd{\mu_0} G(t,x,\lambda|x_0,\lambda_0)G(t,y,\mu|y_0,\mu_0)\expval{\rho(0,x_0,\lambda_0)\rho(0,y_0,\mu_0)}\upd{c}.
\end{align}

The initial state that is typically assumed in GHD is a local equilibrium state, which are states that have quasi-local correlations:
\begin{align}
    \expval{\rho(0,x_0,\lambda_0)\rho(0,y_0,\mu_0)}\upd{c} &= \delta(x_0-y_0) C(x,\lambda_0,\mu_0),
\end{align}
where $C(x,\lambda_0,\mu_0)$ are the GGE correlations at point $x$. Therefore we find
\begin{align}
    \expval{\rho(t,x,\lambda)\rho_{j}(t,y,\mu)}\upd{c} &= \int\dd{x_0}\dd{\lambda_0}\dd{\mu_0} G(t,x,\lambda|x_0,\lambda_0)G(t,y,\mu|y_0,\mu_0)C(x_0,\lambda_0,\mu_0).
\end{align}

From the general theory of integrability we know that the entropy density is given by the thermodynamic Bethe ansatz \cite{yangyang,Zamolodchikov1990,takahashi_1999} (a discussion of its most general form is found in \cite{10.21468/SciPostPhys.6.4.049}):
\begin{align}
    s = -\int\dd{\lambda}\rho_{{\rm s}}(\lambda) \gamma(n(\lambda)).
\end{align}
Here $\gamma(n)$ is a function that depends on the statistics of particles. For instance classical particles we have $\gamma(n) = n\log n -n$, while for quantum particles with fermionic statistics its $\gamma(n) = n\log n + (1-n)\log(1-n)$. In a model where the entropy is determined by $\gamma(n)$ the GGE correlations for the quasi-particle density are given by~\cite{10.21468/SciPostPhys.5.5.054,SciPostPhys.3.6.039}
\begin{align}
    \expval{\rho(x,\lambda),\rho(y,\mu)}\upd{c}\ind{GGE} &= \delta(x-y)\int\dd{\nu} \tfrac{\rho_{{\rm s}}(x,\nu)}{\gamma''(n(x,\nu))} \qty[\delta(\cdot-\nu)]\upd{drT}(x,\lambda)\qty[\delta(\cdot-\nu)]\upd{drT}(x,\mu).
\end{align}
This immediately follows from the fact that correlation functions in GGE are delta-correlated at the Euler scale, and from the  the free energy of the Thermodynamic Bethe ansatz, which gives the correlation matrix \cite{SciPostPhys.3.6.039}. Note that we can split the correlations in a singular and non-singular part and write the initial correlations as:
\begin{align}
    C(x_0,\lambda_0,\mu_0) &= \tfrac{\rho_{{\rm s}}(0,x_0,\lambda_0)}{\gamma''(n(0,x_0,\lambda_0))} \delta(\lambda_0-\mu_0) + {\text{(continuous in } \lambda)}.
\end{align}

Similarly to the discussion of the solution to the linearized Euler equation also $\expval{\rho(t,x,\lambda)\rho(t,y,\mu)}\upd{c}$ will have jumps and delta peaks on top of a continuous background. We will restrict the discussion to study the behaviour of the correlations close to $x \approx y$, which is also physically most interesting as it gives insights into the local state. One can easily convince oneself that only the singular parts of the initial correlations and the evolution kernel contribute to these. We find:
\begin{align}
    &\expval{\rho(t,x,\lambda)\rho(t,y,\mu)}\upd{c}\eval_{x\approx y}\nonumber\\
    &=\int\dd{x_0}\dd{\lambda_0}\delta(x-Y(t,x_0,\lambda_0))\delta(y-Y(t,x_0,\lambda_0))\qty[\delta(\lambda-\lambda_0)]\upd{drT}\qty[\delta(\mu-\lambda_0)]\upd{drT}\tfrac{\rho_{{\rm s}}(0,x_0,\lambda_0)}{\gamma''(n(0,x_0,\lambda_0))}\nonumber\\
    &+\tfrac{1}{2}\int\dd{x_0}\dd{\lambda_0}\sgn(x-Y(t,x_0,\lambda_0))\delta(y-Y(t,x_0,\lambda_0))\qty[\partial_xn(t,x,\lambda)\vu{T}\qty[\delta(\lambda-\lambda_0)]\upd{drT}]\upd{drT}\nonumber\\
    &\times\qty[\delta(\mu-\lambda_0)]\upd{drT}\tfrac{\rho_{{\rm s}}(0,x_0,\lambda_0)}{\gamma''(n(0,x_0,\lambda_0))}\nonumber\\
    &+\tfrac{1}{2}\int\dd{x_0}\dd{\lambda_0}\delta(x-Y(t,x_0,\lambda_0))\sgn(y-Y(t,x_0,\lambda_0))\qty[\delta(\lambda-\lambda_0)]\upd{drT}\nonumber\\
    &\times \qty[\partial_xn(t,x,\mu)\vu{T}\qty[\delta(\mu-\lambda_0)]\upd{drT}]\upd{drT}\tfrac{\rho_{{\rm s}}(0,x_0,\lambda_0)}{\gamma''(n(0,x_0,\lambda_0))} +\mathrm{(continuous)}.
\end{align}
Here the $\cdot\upd{drT}$ acts on $\lambda$ in the first bracket and on $\mu$ in the second bracket. From the definition \eqref{equ:sm_linGHD_Y0} we can infer:
\begin{align}
    \dv{Y(t,x_0,\lambda)}{x_0} = \frac{1}{\dv{Y^0(t,x,\lambda)}{x}}\eval_{x=Y(t,x_0,\lambda)} = \frac{\rho_{{\rm s}}(0,x_0,\lambda)}{\rho_{{\rm s}}(t,Y(t,x_0,\lambda),\lambda)},
\end{align}
which is positive and thus we finally find:
\begin{align}\label{equ:corr_sing}
    &\expval{\rho(t,x,\lambda)\rho(t,y,\mu)}\upd{c}\eval_{x\approx y} =\delta(x-y)\int\dd{\lambda_0}\qty[\delta(\lambda-\lambda_0)]\upd{drT}\qty[\delta(\mu-\lambda_0)]\upd{drT}\tfrac{\rho_{{\rm s}}(t,x,\lambda_0)}{\gamma''(n(t,x,\lambda_0))}\nonumber\\
    &+\tfrac{1}{2}\sgn(x-y)\int\dd{\lambda_0}\qty[\partial_xn(t,x,\lambda)\vu{T}\qty[\delta(\lambda-\lambda_0)]\upd{drT}]\upd{drT}\qty[\delta(\mu-\lambda_0)]\upd{drT}\tfrac{\rho_{{\rm s}}(t,y,\lambda_0)}{\gamma''(n(t,y,\lambda_0))}\nonumber\\
    &+\tfrac{1}{2}\sgn(y-x)\sum_{i_0}\int\dd{\lambda_0}\qty[\delta(\lambda-\lambda_0)]\upd{drT}\qty[\partial_xn(t,x,\mu)\vu{T}\qty[\delta(\mu-\lambda_0)]\upd{drT}]\upd{drT}\tfrac{\rho_{{\rm s}}(t,x,\lambda_0)}{\gamma''(n(t,x,\lambda_0))}\nonumber\\
    &+\mathrm{(continuous)}.
\end{align}
The first term is precisely the delta correlation part from the GGE, as expected, since the state should be in local equilibrium. However, there are more contributions, which correspond to the long range correlations first discovered in~\cite{PhysRevLett.131.027101,10.21468/SciPostPhys.15.4.136}. The interesting result of this investigation is that the long range correlations do not vanish at $x\to y$. In general there is a jump at $x = y$, and also a continuous background. Note also that the discontinuity at $x=y$ is proportional to $\partial_x n(t,x,\lambda)$ (i.e.\ it vanishes in a GGE state) and immediately appears at any (Euler scale) time $t>0$. 

\subsection{Correlations of normal modes}
We can also express the solution in terms of the corresponding perturbation of the occupation function (this gives correlations of normal modes). Note that from the defining relation of $\rho_{{\rm s}}(t,x,\lambda)$ we have:
\begin{align}
     \rho(t,x,\lambda) &= \tfrac{1}{2\pi}n(t,x,\lambda ) + n(t,x,\lambda)\vu{T}\rho(t,x,\lambda).
\end{align}
Expanding up to linear order in perturbation $\delta \rho(x,\lambda)$ we find:
\begin{align}
    \delta \rho(t,x,\lambda) &= \tfrac{1}{2\pi}\delta n(t,x,\lambda )+ \delta n(t,x,\lambda)\vu{T}\rho(t,x,\lambda) + n(t,x,\lambda)\vu{T}\delta \rho(t,x,\lambda)\nonumber\\
    &= 1\upd{dr}(t,x,\lambda)\delta n(t,x,\lambda ) + n(t,x,\lambda)\vu{T}\delta \rho(t,x,\lambda).
\end{align}
From this we find the relation $\delta \rho(t,x,\lambda) = \tfrac{1}{2\pi}(1\upd{dr}(t,x,\lambda) \delta n(t,x,\lambda))\upd{drT} = (\rho_{{\rm s}}(t,x,\lambda)\delta n(x,\lambda))\upd{drT}$. Therefore we have:
\begin{align}
    \expval{\delta n(t,x,\lambda)\delta n(s,y,\mu)} &= \frac{1}{\rho_{{\rm s}}(x,\lambda)\rho_{{\rm s}}(y,\mu)}(\vb{1}-n(t,x,\lambda)\vu{T})(\vb{1}-n(s,x,\mu)\vu{T})\expval{\rho(t,x,\lambda)\rho(s,y,\mu)}\upd{c}.
\end{align}
Inserting \eqref{equ:corr_sing} into this we can read off the singular part of the occupation function correlations:
\begin{align}
    &\expval{\delta n(t,x,\lambda)\delta n(t,y,\mu)}\eval_{x\approx y} =\delta(x-y)\delta(\lambda-\mu)\tfrac{1}{\rho_{{\rm s}}(x,\lambda)\gamma''(n(t,x,\lambda))}\nonumber\\
    &+\tfrac{1}{2}\sgn(x-y)\partial_xn(t,x,\lambda)\qty(\vu{T}\qty[\delta(\cdot-\mu)]\upd{drT})\eval_{\cdot = \lambda}\tfrac{1}{\rho_{{\rm s}}(x,\lambda)\gamma''(n(t,x,\mu))}\nonumber\\
    &+\tfrac{1}{2}\sgn(y-x)\partial_xn(t,x,\mu)\qty(\vu{T}\qty[\delta(\cdot-\lambda)]\upd{drT})\eval_{\cdot = \mu}\tfrac{1}{\rho_{{\rm s}}(x,\mu)\gamma''(n(t,x,\lambda))}\nonumber\\
    &+\mathrm{(continuous)}.
\end{align}
Again the $\delta$ peak corresponds to the GGE result~\cite{10.21468/SciPostPhys.5.5.054,PhysRevB.54.10845}:
\begin{align}
    \expval{\delta n(x,\lambda)\delta n(y,\mu)}\ind{GGE} &= \delta(x-y)\delta(\lambda-\mu)\tfrac{1}{\rho_{{\rm s}}(x,\lambda)\gamma''(n(x,\lambda))}.
\end{align}

\subsection{Direct solution to the linear-response evolution equation}

Instead of using a linear response analysis of the fixed-point problem in order to determine the linear-response kernel (and thus the two-point correlation functions), one can directly solve the inhomogeneous evolution equation by using height fields. In order to make explicit the level of generality taken here, we use the formal Euler-hydrodynamic notation, where an index $I$ parametrises conserved quantities; it corresponds, in GHD, to $I=\lambda$. We use the flux Jacobian $A_I^{~J}$ and correlation matrix $C_{IJ}$ (corresponding to that discussed above), as well as the diagonalisation matrix $R_I^{~K}$; these concepts and their implementation in GHD are explained in \cite{10.21468/SciPostPhysLectNotes.18}.

From \cite{10.21468/SciPostPhys.5.5.054}, a linear perturbation on top of an inhomogeneous background evolves as
\begin{equation}\label{rholinear}
    \partial_t \delta\rho_I(t,x)
    + \partial_x (A_I^{~K}(t,x)\delta\rho_K(t,x)) = 0.
\end{equation}
This can be written in quasi-linear form by using the (spatially symmetrised) height field $\delta N_I(t,x) = \frac12\int_{-\infty}^\infty \dd{x'}\sgn(x-x')\delta\rho_I(t,x')$,
\begin{equation}
    \partial_t \delta N_I(t,x)
    + A_I^{~J}(t,x)\partial_x \delta N_J(t,x) = 0.
\end{equation}

The flux Jacobian is diagonalised as
\begin{equation}\label{Aveff}
    A = R^{-1} v^{\rm eff} R
\end{equation}
where $v^{\rm eff}$ is the diagonal matrix of hydrodynamic velocities $v^{\rm eff}_I$. In GHD,
\begin{equation}
    R_I^{~J} = (1-n\vu{T})_I^{~J}
\end{equation}
and therefore $R^{-1}$ implements the transposed dressing $\cdot\upd{drT}$ discussed above. In GHD, the $R$ matrix satisfies itself an evolution equation, which can be written in two forms,
\begin{equation}
    \label{Revol}
    \partial_t R + v^{\rm eff}\partial_x R = 0,\quad
    \partial_t R^{-1} + A \partial_x R^{-1} = 0.
\end{equation}
This was noticed and used in \cite{10.21468/SciPostPhys.15.4.136}. Using this, one observes that $\delta \hat N_I(t,x) = R_I^{~J}(t,x) \delta N_J(t,x)$ diagonalises the evolution equation,
\begin{equation}
    \partial_t \delta \hat N_I(t,x)
    + v^{\rm eff}_I(t,x) \partial_x\delta\hat N_I(t,x) = 0.
\end{equation}
The solution is immediate in terms of the characteristic function $u_I(t,x)$,
\begin{equation}
    \delta \hat N_I(t,x) = \delta \hat N_I(0,u_I(t,x)).
\end{equation}
Therefore
\begin{equation}
    \delta N_I(t,x)
    =(R^{-1})_I^{~J}(t,x)
    R_J^{~K}(0,u_J(t,x))
    \delta N_K(0,u_J(t,x))
\end{equation}
which, by taking the $x$ derivative, reproduces, in GHD, the kernel Eq.~13 in the main text. We note that, as per \eqref{Revol}, we have $R_J^{~K}(0,u_J(t,x)) = R_J^{~K}(t,x)$. In particular, the two-point function of conserved densities $\rho_M(t,x)$ in an initial delta-correlated state
\begin{equation}
    \expval{\rho_M(x,0)\rho_N(y,0)}^{\rm c}
    = \delta(x-y)C_{MN}(x)
\end{equation}
is, according to the principles of the BMFT,
\begin{equation}\begin{aligned}
    &\expval{\rho_M(t,x)\rho_N(s,y)}^{\rm c}
    =\\ &
    -\partial_x\partial_y \Bigg(
    (R^{-1})_M^{~I}(t,x)R_I^{~K}(0,x)
    \int_{u_J(s,y)}^{u_I(t,x)} \dd x'\,C_{KL}(x')
    \,\Theta(u_I(t,x)-u_J(s,y))
    R_J^{~L}(0,y)(R^{-1})_N^{~J}(s,y)
    \Bigg).
    \end{aligned}
\end{equation}

This is a particularly simple form of the full two-point function, including long-range correlations. Eq.~\eqref{rholinear} is expected to hold for any Euler-scale hydrodynamic system, and in general the flux Jacobian is diagonalisable as per \eqref{Aveff}. Here, the only structure of GHD used is the evolution equation \eqref{Revol} for the $R$ matrix. Thus, this is a general result for any hydrodynamic system with this structure.

\subsection{Simplified formulas for the hard rods system}
In the special case of the hard rods system with length $d$ (one particle species with $v(\lambda) = \lambda$ and $\varphi(\lambda,\mu) = -d$) the formulas simplify: Note that $\cdot\upd{dr}$ and $\cdot\upd{drT}$ can be explicitly evaluated and are given by~\cite{Doyon_2017,SciPostPhys.3.6.039}:
\begin{align}
    f\upd{dr}(t,x,\lambda) &= f(t,x,\lambda) - d \int\dd{\mu} \rho(t,x,\mu)f(t,x,\mu) &  f\upd{drT}(t,x,\lambda) &= f(t,x,\lambda) - d \rho(t,x,\lambda) \int\dd{\mu} f(t,x,\mu).
\end{align}
From this one can also obtain the following explicit formulas:
\begin{align}
    n(t,x,\lambda) &= \frac{2\pi\rho(t,x,\lambda)}{1-a\int\dd{\mu} \rho(t,x,\mu)} & \rho(t,x,\lambda) &= \frac{1}{2\pi}\frac{n(t,x,\lambda)}{1+\tfrac{d}{2\pi}\int\dd{\mu}n(t,x,\mu)}.
\end{align}
Using this we can explicitly write \eqref{equ:linGHD_kernel}:
\begin{align}
    G(t,x,\lambda|x_0,\lambda_0) &= \partial_x\Big[\sgn(x-Y(t,x_0,\lambda_0))(\delta(\lambda-\lambda_0) - d \rho(t,x,\lambda)) - \sgn(x-Y(t,x_0,\lambda))\tfrac{d\rho(t,x,\lambda)}{1-d\int\dd{\mu}\rho(t,x,\mu)}\nonumber\\
    &\indent+ \tfrac{d^2\rho(t,x,\lambda)}{1-d\int\dd{\mu}\rho(t,x,\mu)}\int\dd{\mu}\sgn(x-Y(t,x_0,\mu))\rho(t,x,\mu)\Big].
\end{align}
Again, inserting this into \eqref{equ:corr_equal_time} gives an explicit formula for the equal time correlations after some time $t$.





\end{document}